\documentclass[twocolumn,superscriptaddress,showpacs,prd,aps,amsmath,amssymb,nofootinbib]{revtex4-1}
\usepackage{natbib}
\usepackage{graphicx,color}
\usepackage{amsmath,amssymb}
\usepackage{verbatim}
\usepackage{float}
\usepackage{wasysym}
\usepackage{amssymb,graphicx}
\usepackage{epsfig}
\usepackage{psfrag}
\usepackage{dsfont}
\usepackage{amsfonts}
\usepackage{mathrsfs}
\usepackage{multirow}
\usepackage{times}
\usepackage{bm}
\usepackage{hyperref}
\hypersetup{
  colorlinks=true,        
  linkcolor=blue,         
  citecolor=cyan,         
}
\begin{document}

\title{Jet launching from binary black hole-neutron star mergers:
  Dependence on black hole spin, binary mass ratio, and  magnetic field
  orientation}
\author{Milton Ruiz}
\affiliation{Department of Physics, University of Illinois at
  Urbana-Champaign, Urbana, IL 61801}
\author{Stuart L. Shapiro}
\affiliation{Department of Physics, University of Illinois at
  Urbana-Champaign, Urbana, IL 61801}
\affiliation{Department of Astronomy \& NCSA, University of
  Illinois at Urbana-Champaign, Urbana, IL 61801}
\author{Antonios Tsokaros}
\affiliation{Department of Physics, University of Illinois at
    Urbana-Champaign, Urbana, IL 61801}

\begin{abstract}
  Black hole-neutron star (BHNS) mergers are one of the most promising targets for multimessenger
  astronomy. Using  general relativistic magnetohydrodynamic simulations of BHNS undergoing
  merger we previously showed that a magnetically--driven jet can be launched by the disk +
  spinning black hole remnant {\it if} the neutron star is endowed with a dipole magnetic field
  extending from the interior into the exterior as in a radio pulsar. These self-consistent studies
  considered a BHNS system with mass ratio $q=3:1$, black hole spin $a/M_{\rm BH}=0.75$ aligned with
  the total orbital angular momentum, and a neutron star that is irrotational, threaded by an aligned
  magnetic field, and modeled by an $\Gamma$--law equation of state with $\Gamma=2$. Here,
  as a crucial step in establishing BHNS systems  as viable progenitors of  central engines that
  power short gamma--ray bursts (sGRBs) and thereby solidify their role as multimessenger sources,
  we survey different BHNS configurations that differ in the spin of the BH companion ($a/M_{\rm BH}
  =-0.5,\,0,\,0.5,\,0.75$), in the mass ratio ($q=3:1$ and $q=5:1$), and in the orientation of the
  magnetic field (aligned and tilted by $90^\circ$ with respect to the orbital angular momentum). We
  find that by $\Delta t\sim 3500M-4000M \sim 88(M_{\rm NS}/1.4M_\odot){\rm ms}-100(M_{\rm NS}/
  1.4M_\odot)\rm ms$ after the peak gravitational wave signal a magnetically--driven jet is launched
  in the cases where the initial spin of the BH companion is $a/M_{\rm BH}= 0.5$ or $0.75$.  The
  lifetime of the jets [$\Delta t\sim 0.5(M_{\rm NS}/1.4M_\odot){\rm s}-0.7(M_{\rm NS}/1.4M_\odot)
    \rm s$] and their outgoing Poynting luminosities [$L_{jet}\sim 10^{51\pm 1}\rm erg/s$] are
  consistent with typical sGRBs, as well as with the Blandford--Znajek mechanism  for launching
  jets and their associated Poynting luminosities. By the time we terminate our simulations,  we
  do not observe either an outflow or a large-scale magnetic field collimation in the other
  configurations we simulate. These results suggest that future multimessenger detections from BHNSs
  are more likely produced by binaries with  highly spinning BH companions and small tilt-angle
  magnetic fields, though other physical processes do not considered here, such as neutrino
  annihilation, may help to power jets in general cases.
\end{abstract}

\pacs{04.25.D-, 04.25.dg, 47.75.+f}
\maketitle

\section{Introduction}
The era of multimessenger astronomy has accelerated with the detection of 
GW170817~\cite{TheLIGOScientific:2017qsa}, a gravitational wave (GW)
signal from the coalescence of a compact binary,
accompanied by  electromagnetic (EM) counterpart radiation across the EM
spectrum~(see e.g.~\cite{GBM:2017lvd,Monitor:2017mdv,Abbott:2017wuw,Chornock:2017sdf,
  Cowperthwaite:2017dyu,Kasen:2017sxr,Nicholl:2017ahq} and reference therein). From
the gravitational radiation signal alone, the inferred  masses of the individual binary
companions are in the broad range of $0.86-2.26\, M_\odot$, though the total mass of
the system is constrained to be~$2.73-3.29\, M_\odot$ with $90\%$ confidence
\cite{TheLIGOScientific:2017qsa}. These estimates, along with the EM counterparts,
and, in particular, the detection of a short gamma--ray burst (sGRB) -- GRB 170817A--
$1.7$s--following the inferred merger time by the Fermi Gamma-Ray Burst Monitor
\cite{FERMI2017GCN} and INTEGRAL~\cite{Savchenko:2017ffs,Savchenko17GCN}, as well as
the associated kilonova/macronova, demonstrate the presence of matter~\cite{GBM:2017lvd}.
These observations strongly suggest a merging binary neutron star system (NSNS) as the
source of GW170817, although they cannot rule out the possibility that one of the binary
companions is a stellar--mass black hole (BH).
Recently, a summary of possible low--mass BH formation channels, and routes by which they
may arise in binaries with a NS companion, have been presented in~\cite{Yang:2017gfb}.

Due to the limited sensitivity of the  current LIGO/Virgo GW laser interferometers there
is still an open question regarding the nature of the GW170817  remnant if one assumes
that its progenitor is a NSNS system (see e.g.~\cite{2017arXiv171007579S,Piro:2018bpl,
  2018ApJ...860...57A,2018ApJ...861..114Y,2018ApJ...861L..12L,Margalit:2017dij,Ruiz:2017due}).
Using EM constraints on the remnant
imposed by the kilonova observations~\cite{FERMI2017GCN,Savchenko:2017ffs,Savchenko17GCN,
  GBM:2017lvd} along with the GW data, it was argued in~\cite{Margalit:2017dij}
that the GW170817 NSNS remnant resulted in a hypermassive NS (HMNS) undergoing collapse
to a BH in $\sim 10^{-2}$s$-1$s. This  hypothesis was supported by our GRMHD simulations
reported in~\cite{Ruiz:2017due}~where we showed that a long-lived,  HMNS seeded with a
pulsar-like magnetic field does not power magnetically-driven and sustained outflows (jets)
believed to be crucial for generating GRBs as in GW170817.
The astrophysical implication of these observations  create therefore the urgent need to
model GWs and EM counterparts from both NSNS and BHNS systems~\cite{Hinderer:2018pei}.

GW170718 and GRB 170817A  provide the best direct confirmation so far that the
merger of compact binaries in which at least one NS is involved can be the engine
that powers sGRBs. This identification was originally proposed by~\cite{Pac86ApJ,EiLiPiSc,NaPaPi}
and recently demonstrated by self-consistent simulations in full general relativistic
magnetohydrodynamics (GRMHD) of  merging BHNSs~\cite{prs15} and merging NSNSs that
undergo {\it delayed} collapse~\cite{Ruiz:2016rai}. The numerical studies in~\cite{prs15}
(hereafter Paper I), whose initial configuration is a BHNS binary with mass ratio $q=3:1$
in a quasicircular orbit, with an NS modeled as an irrotational $\Gamma=2$ polytrope and
a BH with dimensionless spin $\tilde{a}\equiv a/M_{\rm BH}=0.75$, showed that a collimated,
mildly relativistic outflow --an incipient jet-- can be launched from the highly spinning
BH remnant surrounded by a magnetized accretion disk. Such a jet requires that a strong poloidal
magnetic field component which connects the disk to the BH poles persist after the disruption
of the NS~\cite{GRMHD_Jets_Req_Strong_Pol_fields,UIUC_PAPER2}. This key feature was achieved
in Paper I by seeding the NS initially with a dipole magnetic field that extends from the
stellar interior into the exterior in a pulsar-like, force-free exterior magnetosphere
(see e.g.~\cite{Ruiz:2014zta}). Following the onset of tidal disruption,
it was found that magnetic winding and the magnetorotational instability (MRI) amplify
the magnetic field above the BH poles from $\sim 10^{13}(1.4M_\odot/M_{\rm NS})$G
when the disk first settles to
$\sim 10^{15}(1.4M_\odot/M_{\rm NS})$G, and this field  eventually drives and confines
the incipient jet by $\Delta t\sim 100(M_{\rm NS}/1.4M_\odot)$ms after peak GW emission.
The lifetime of the jet and the outgoing
Poynting luminosity are~$\Delta t\sim 0.5(M_{\rm NS}/1.4M_\odot)$s and
$L_{\rm EM}\sim 10^{51}\rm erg/s$, values which are both  consistent with
typical sGRBs~(see e.g.~\cite{Bhat:2016odd,Lien:2016zny,Svinkin:2016fho}).

In the NSNS scenario, by contrast, an incipient jet emerges whether or not the initial
poloidal magnetic field is confined to the NS interior, as long as the binary forms a
HMNS that undergoes delayed collapse to a BH~\cite{Ruiz:2016rai}.
During the formation and spindown of the transient, differentially-rotating HMNS magnetic winding and
both the Kelvin-Helmholtz instability and the MRI boost the rms value of the magnetic field
to~$\gtrsim~10^{15.5}$G~\cite{Kiuchi:2014hja,Kiuchi:2015sga}.  In the prompt collapse
scenario, the~onset of BH formation following the NSNS merger prevents that amplification
\cite{Ruiz:2017inq}.  The calculations in~\cite{Ruiz:2016rai} that model the NS with a
simple $\Gamma$--law equation of state (EOS) with $\Gamma=2$, allowing for shock heating,
show that the disk + BH remnant launches a jet at about
$\sim 44(M_{\rm NS}/1.8M_\odot)\rm ms$ following the NSNS merger, which lasts $\Delta
t\sim 97(M_{\rm NS}/1.8M_\odot)$ms. The outgoing Poynting luminosity is~$L_{\rm EM}\sim
10^{51}\rm erg/s$, consistent with short sGRBs~(see e.g.~\cite{Bhat:2016odd,
  Lien:2016zny,Svinkin:2016fho}). Recent GRMHD simulations of NSNS mergers reported
in~\cite{Kawamura:2016nmk,Ciolfi:2017uak}, in which the effects of different EOSs, different
mass ratios, and different magnetic field orientations with an initial strength of
$\sim 10^{12}\rm G$ were studied, did not find evidence of
an outflow or a jet after $\Delta t\sim 35\rm ms$ following the NSNS merger, although the formation
of an organized magnetic field structure above the BH was  observed. A lack of a jet in the
high resolution NSNS mergers has been also reported~\cite{Kiuchi:2014hja}, in which the NS
is modeled by an H4 EOS. At the end of those simulations, however, they report persistent
fall-back debris in the atmosphere, which increases the ram pressure  above the BH poles,
preventing the system form approaching a near force-free environment as required  for jet
launching. A longer time integration may be needed for the atmosphere to disperse and for
the jet to emerge. Note that jet
launching may not be possible for all EOSs, if the matter fall-back timescale is longer
than the disk accretion timescale~\cite{Paschalidis:2016agf}. The seeded poloidal magnetic
field  in the numerical studies of~\cite{Kawamura:2016nmk,Ciolfi:2017uak,Kiuchi:2014hja}
is restricted to the NS interior.

In this paper, we survey fully relativistic BHNS configurations initially in a quasicircular orbit
that undergo merger to address the question: {\it Can all the BHNS configurations that
  undergo merger in which the NS is seeded with a pulsar-like, force-free magnetic field be
  progenitors of the engine that launches incipient jets?}

In particular, we now consider BHNS configurations with mass ratio $q=3:1$ in which the dimensionless
spin of the BH companion is $\tilde{a}=-0.5$ (counter--rotating), $\tilde{a}=0$ (nonspinning),
and $\tilde{a}=\,0.5$, all aligned with the orbital angular momentum. In addition, we consider a BHNS
configuration with mass ratio $q=5:1$ in  which the BH companion has no spin initially. In all
cases, the NS is endowed with a dynamically weak poloidal magnetic field that extends
from the stellar interior into the  NS  exterior (i.e. a pulsar-like magnetic field) whose dipole
magnetic moment is also aligned with the orbital angular momentum.
Finally, to study the effect of different magnetic field topologies on the jet launching,
we evolve the same configuration as in Paper I (mass ratio $q=3:1$ and BH spin $\tilde{a}=0.75$)
but now seed the NS with a pulsar-like magnetic field whose dipole magnetic moment is
tilted $90^o$ with respect to the orbital angular momentum. Following Paper I, we model the
initial stars as irrotational $\Gamma=2$ polytropes.

In agreement with our earlier calculations, where the star is seeded with a dipole magnetic
field confined to the stellar interior~\cite{Etienne:2007jg,Etienne:2011ea}, we find that the
BHNS mergers listed above lead to a disk + BH remnant with a rest-mass ranging from $\sim 10^{-3}
M_{\odot}(k/189.96\rm km^2)^{1/2}$ to $\sim 10^{-1}M_{\odot}(k/189.96\rm km^2)^{1/2}$, and
dimensionless spin ranging from $\tilde{a}\sim 0.33$ to  $\sim 0.85$. Here $k$ is the polytropic
gas constant defined as $k=P/\rho_0^\Gamma$, where $P$ and $\rho_0$ are the {\it initial} cold pressure
and the rest-mass density (see below). The early evolution, tidal disruption and the merger phases
are  unaltered by the dynamically weak initial magnetic field. In the post-merger phase we find that,
as in Paper I, by around $\Delta t\sim 3500M \approx 88(M_{\rm NS}/1.4M_\odot)\rm ms$  after the GW
peak emission a magnetically--driven jet
is launched in the case where the initial spin of the BH companion is $\tilde{a}=0.5$.  The
lifetime of the jet [$\Delta t\sim 0.7(M_{\rm NS}/1.4M_\odot)\rm s$] and outgoing Poynting luminosity
[$L_{jet}\sim 10^{52}\rm erg/s$] are consistent with observations of sGRBs (see e.g.
\cite{Bhat:2016odd}),~as well as with the Blandford--Znajek (BZ)~\cite{BZeffect} mechanism
for launching jets and their associated Poynting luminosities~\cite{MembraneParadigm}.
In contrast, by the time we terminate our simulations, we do not find any indication of an
outflow in the other cases; in the nonspinning case ($\tilde{a}=0$), where a persistent
fall-back debris toward the BH is observed until the end of the simulation, the magnetic
field above the BH poles is wound into a helical configuration, but the magnetic pressure gradients are 
still too weak to overcome the fall-back ram pressure, and thus it is expected that a longer simulation
is required if a  jet were to emerge. However, if the fall-back debris timescale is longer
than the disk accretion timescale [$\Delta t\sim 0.36(M_{\rm NS}/1.4M_\odot)\rm s$], the jet launching
in this case may be suppressed. By contrast, in the counter rotating BHNS configuration
the star plunges quickly into the BH, leaving an ``orphan'' BH with a negligibly small accretion
disk containing less than $1\%$~of the rest-mass of the NS. Similar behavior is observed in the
BHNS configuration with mass ratio $q=5:1$. Finally, in the tilted magnetic field case,
we do not find a coherent poloidal magnetic field component remaining after the BHNS merger,
hence the key~ingredient for jet launching~\cite{GRMHD_Jets_Req_Strong_Pol_fields} is absent.

These preliminary results suggest that jet launching  may strongly depend on a threshold value of
(a) the initial black hole spin, which, along with the tidal-break up separation, controls
the mass of the accretion disk, and (b) the tilt-angle of the magnetic field, which triggers
the presence of a poloidal component of the magnetic field in the post-merger
phase. So future multimessenger detections from BHNSs are most likely produced by binaries
with a highly--spinning BH companion and small tilt-angle magnetic fields (see
also~\cite{Bhattacharya:2018lmw}).

The remainder of the paper is organized as follows: A short summary of our numerical
methods and their implementation is presented in Sec.~\ref{subsec:evolution_code}. A
detailed description of our adopted initial data and the grid structure  used for solving
the GRMHD equations is given in Sec.~\ref{subsec:idata} and Sec.~\ref{subsec:grid},
respectively.  In Sec.~\ref{subsec:diagnostics}~we describe the diagnostics employed to
monitor and verify the reliability of our numerical calculations. We present our results
in Sec.~\ref{sec:results}, along with a comparison with the results of Paper I, as well as
with the "universal'' analytic model presented in~\cite{Shapiro:2017cny}. Finally, we  offer
conclusions in Sec.~\ref{sec:conclusion}. We adopt geometrized units ($G=c=1$) throughout
the paper except where stated explicitly. Greek indices denote all four spacetime dimensions,
while Latin indices imply spatial parts only.
%
\begin{figure*}[th]
  \centering
  \includegraphics[width=0.49\textwidth]{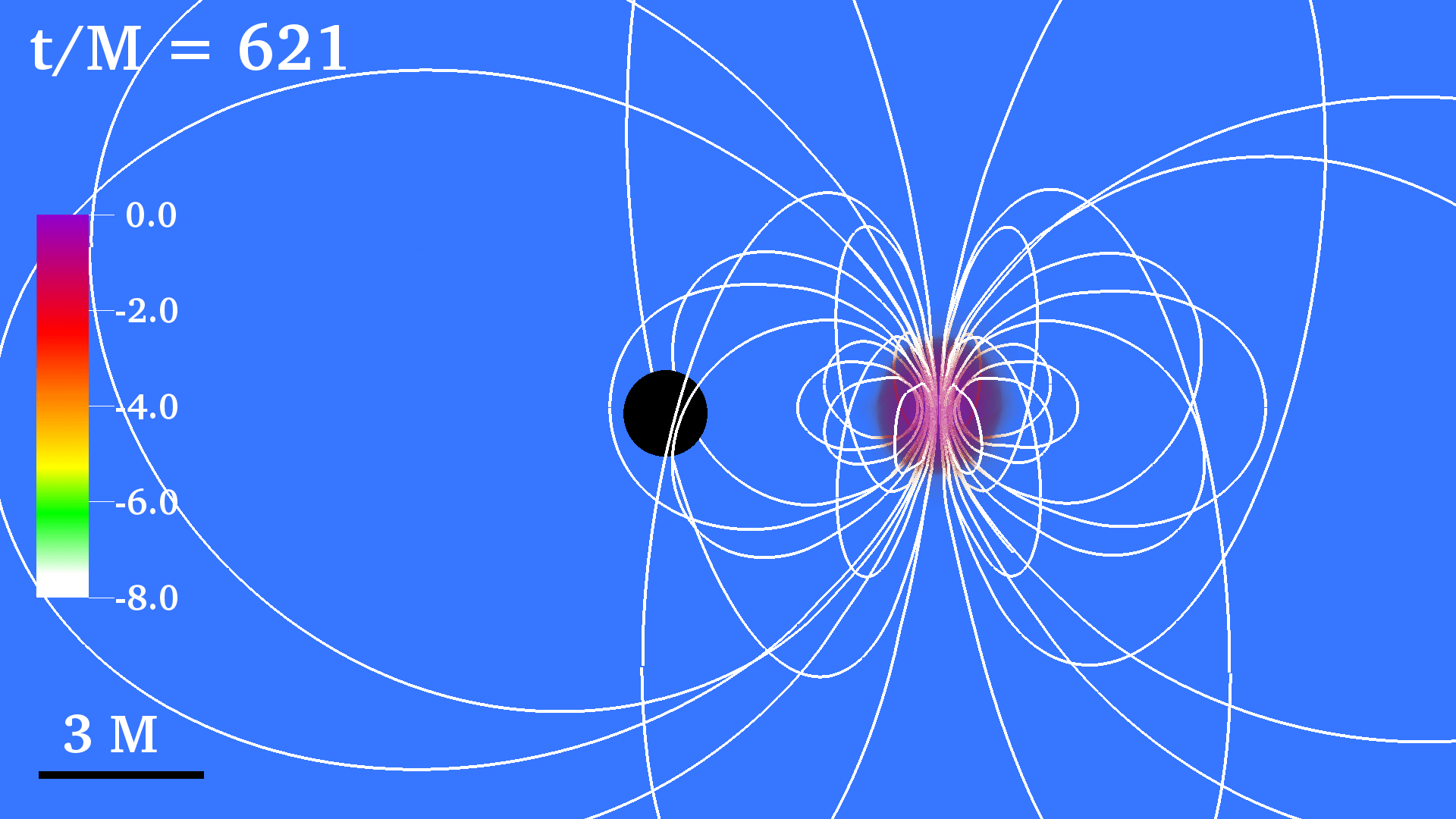}
  \includegraphics[width=0.49\textwidth]{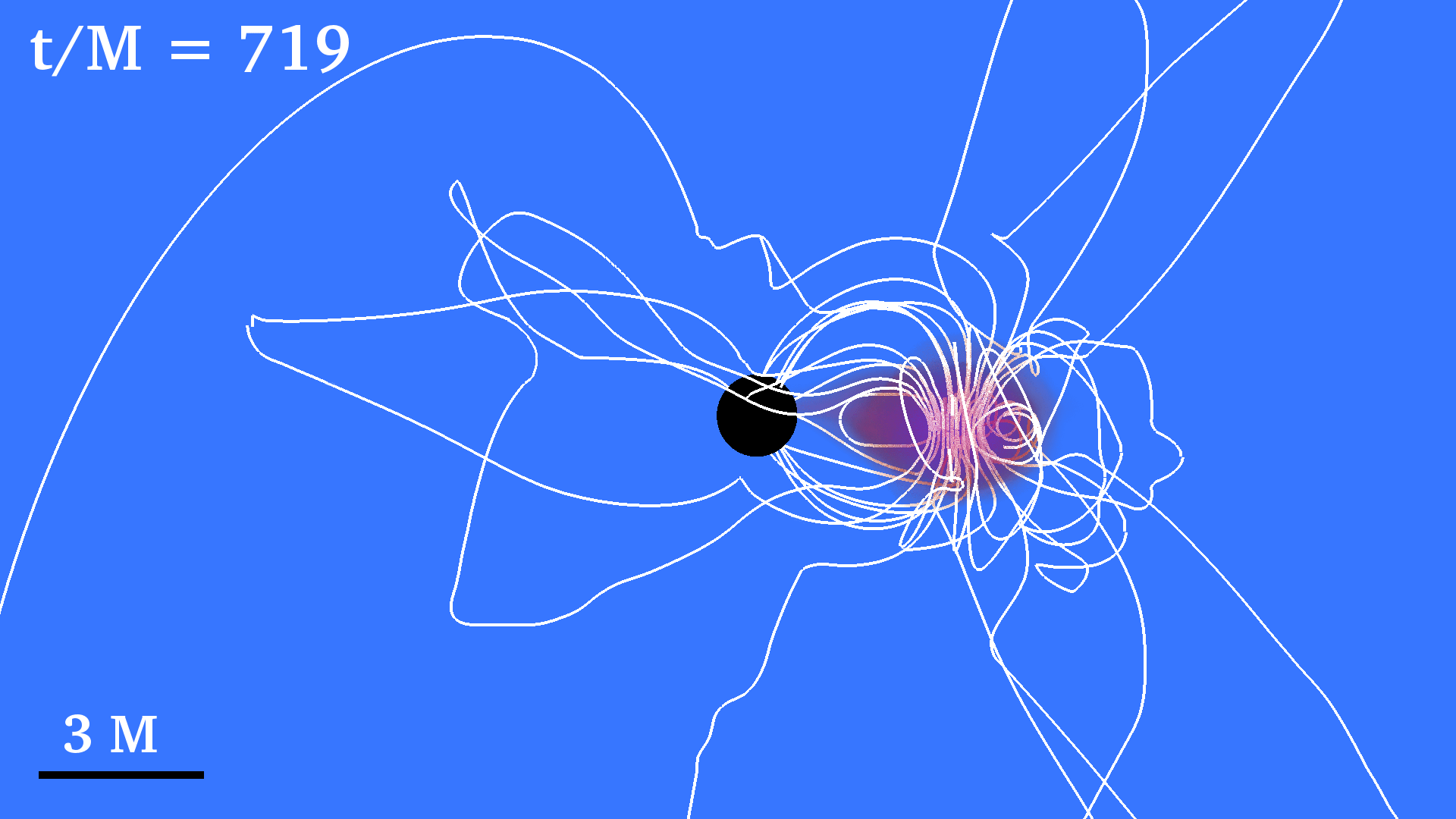}
  \includegraphics[width=0.49\textwidth]{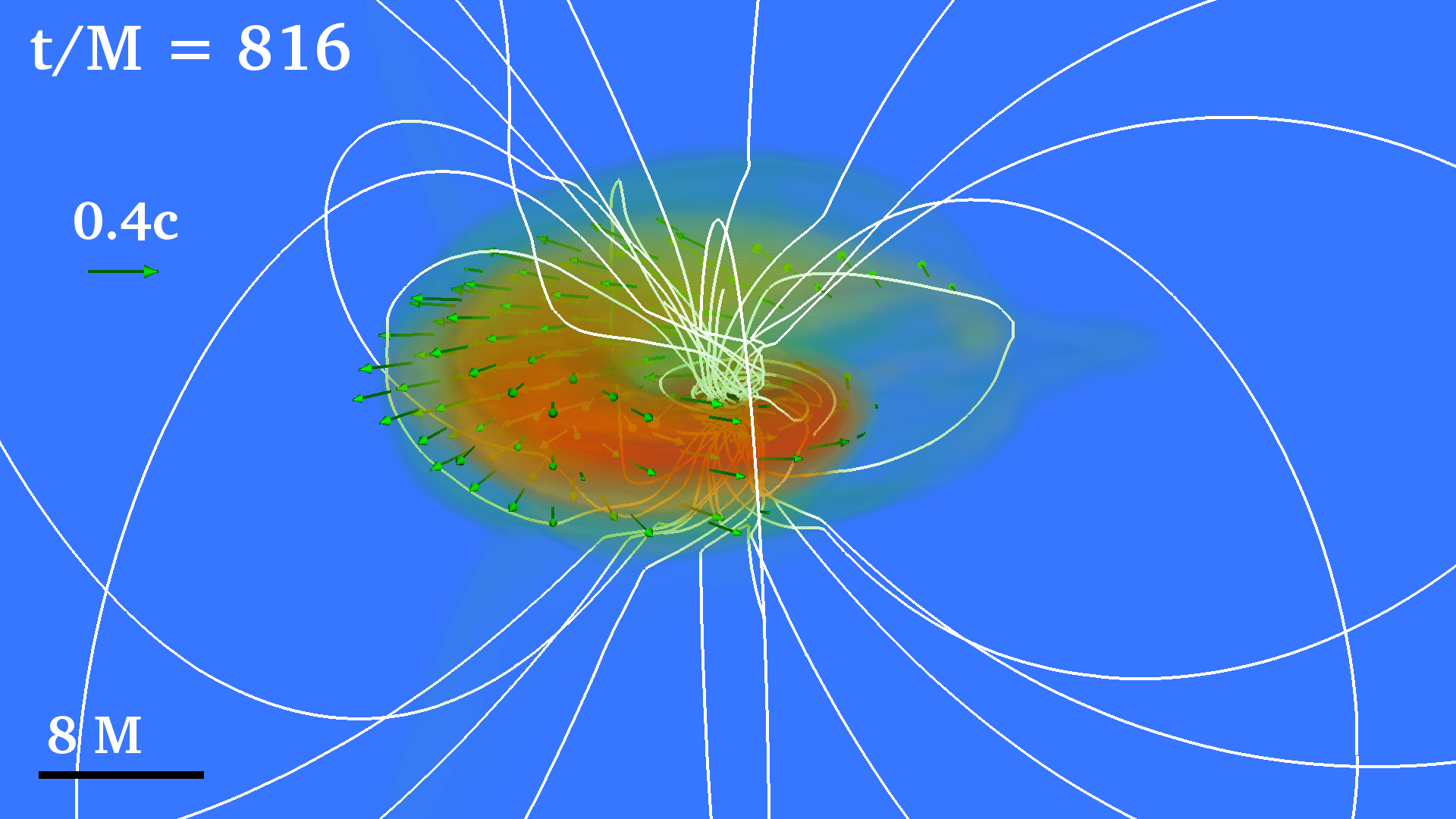}
  \includegraphics[width=0.49\textwidth]{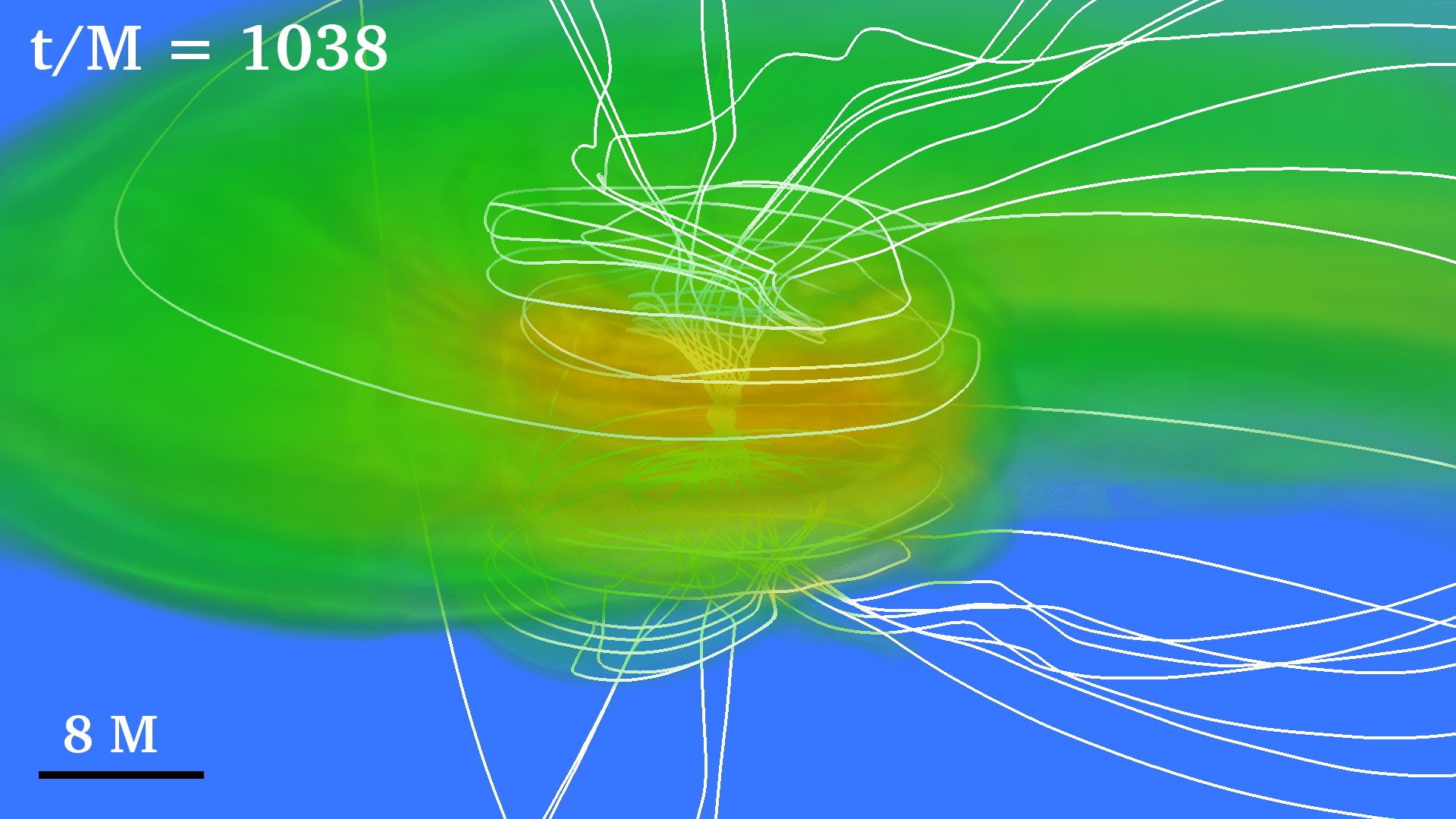}
  \includegraphics[width=0.49\textwidth]{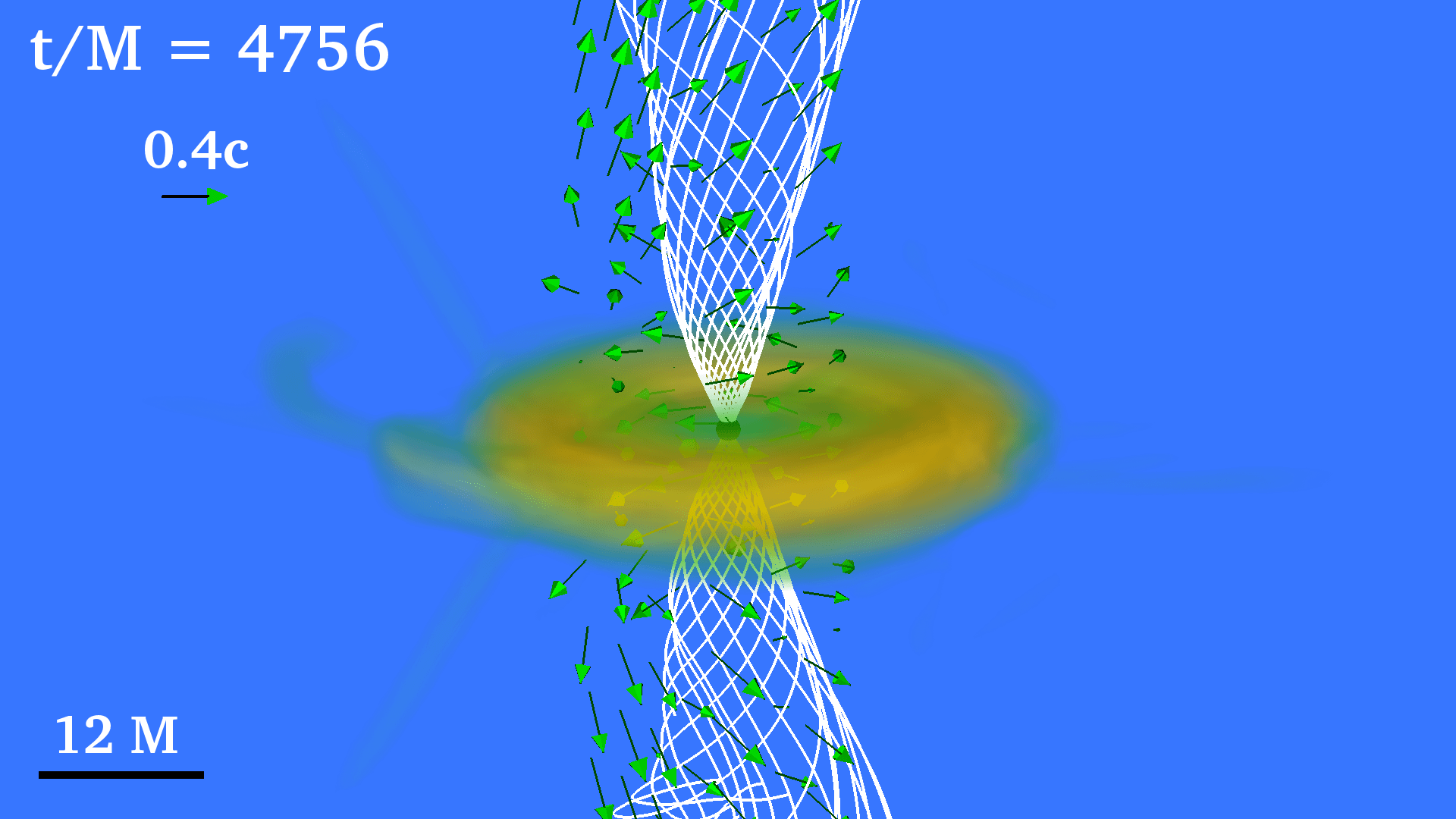}
  \includegraphics[width=0.49\textwidth]{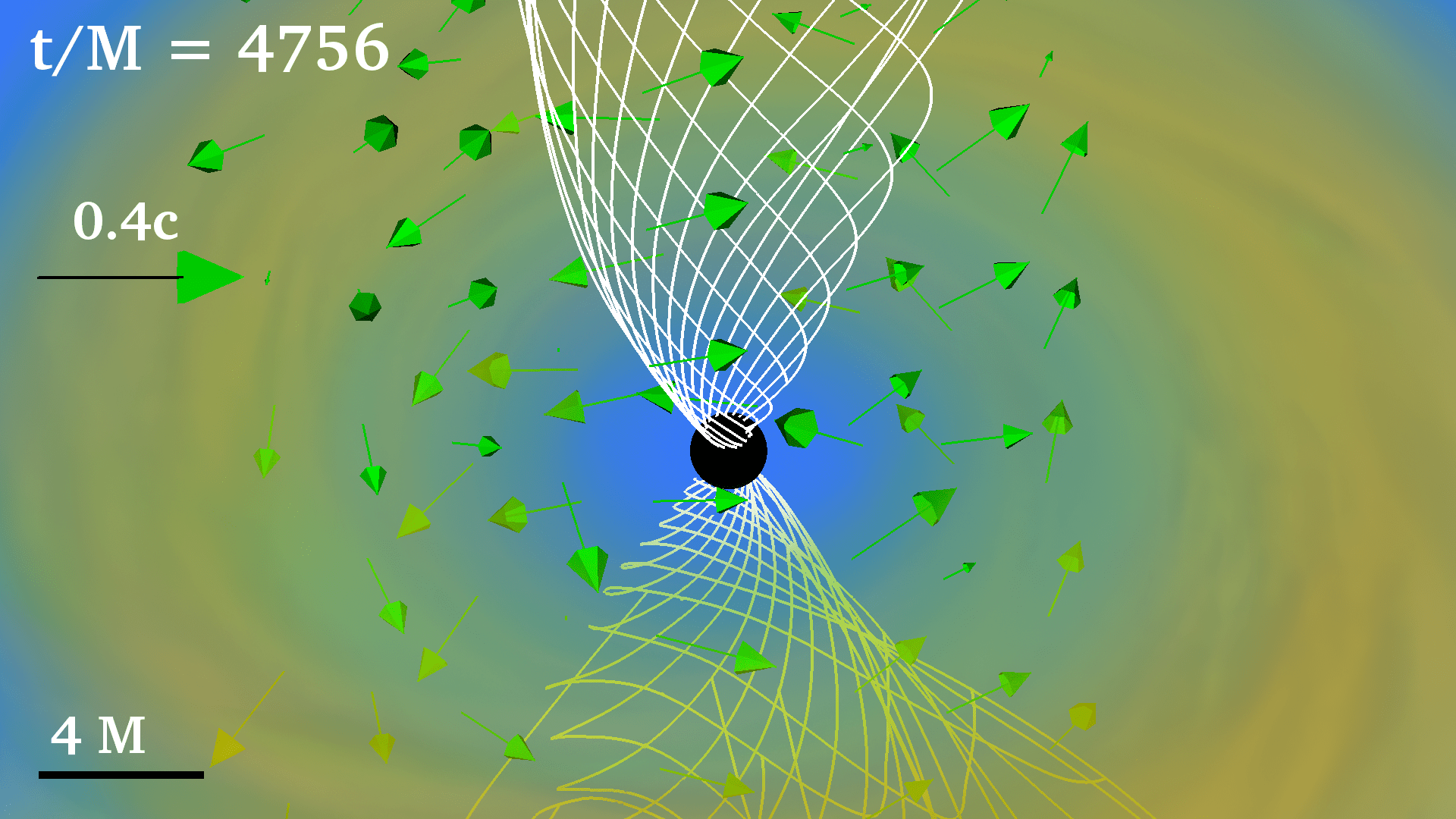}
\caption{Volume rendering of rest-mass density $\rho_0$ normalized to the initial NS
    maximum value $\rho_0=8.92\times 10^{14}\,(1.4M_\odot/M_{\rm NS})^2\rm{g\,/cm}^{3}$
    (log scale) at selected times for case $\rm Aliq3sp0.5$ (see Table~\ref{table:BHNS_ID}).
    White lines denote the magnetic field while the arrows denote the fluid velocity. The
    BH apparent horizon is shown as a black sphere.  Here $M=2.5\times 10^{-2}(M_{\rm NS}/1.4M_\odot)\rm ms$
    = $7.58(M_{\rm NS}/1.4M_\odot)\rm km$.
    \label{fig:BHNS_case_q_3_1_s05}}
\end{figure*}
%
\section{Numerical Methods}
\label{sec:Methods}

The formulation and numerical schemes for  BHNS evolutions have been described in detail
previously in~\cite{Etienne:2007jg,Etienne:2011ea,Etienne:2011re,prs15} and we refer the reader to
those references for further details. In this section we introduce our notation and
briefly summarize our numerical methods.
%
\subsection{Basic Equations}
\label{subsec:evolution_code}
We carry out the numerical evolution using the Illinois GRMHD moving mesh
refinement code that has been embedded in the \texttt{Cactus/Carpet} infrastructure
\cite{AllAngFos01,cactusweb,Carpet,carpetweb}. The code has been tested, including
resolution studies, and used in the past in multiple GRMHD studies involving compact
objects including magnetized BHNS binaries; see e.g.~\cite{Etienne:2012te,UIUC_PAPER1,
  UIUC_PAPER2,prs15}. The code has the following sectors:
%
\paragraph*{\bf Spacetime evolution:}
We  use the $3+1$  formalism of  GR and decompose the full metric of the spacetime
$g_{\mu\nu}$ according to
\begin{align}
  ds^2&=g_{\mu\nu}\,dx^\mu\,dx^\nu\nonumber\\
  &=-\alpha^2\,dt^2+\gamma_{ij}\,\left(dx^i+\beta^i\,dt\right)
  \left(dx^j+\beta^j\,dt\right)\,,
\end{align}
with $\alpha$ and  $\beta^i$ the gauge variables, and $\gamma_{\mu\nu}= g_{\mu\nu}
+n_\mu\,n_\nu$ the three-metric induced on a spatial hypersurface with a timelike
future pointing unit vector $n^\mu=(1/\alpha,-\beta^i/\alpha)$.  Associated
with the time slice we define  the extrinsic curvature $K_{\mu\nu}\equiv-\gamma_{\mu\alpha}
\nabla^\alpha\,n_\nu$. The  spatial three-metric and extrinsic curvature are then evolved
via the Baumgarte--Shapiro--Shibata--Nakamura (BSSN) formulation~\cite{shibnak95,BS}; see
also~\cite{BSBook} for discussion. The dynamical variables are then: (a) the conformal
exponent~$\phi=\rm{ln}(\gamma)/12$, where $\gamma$ is the determinant of the three-metric,
(b) the conformal metric $\tilde{\gamma}_{ij}=e^{-4\phi}\gamma_{ij}$, (c)  the conformal, trace-free
extrinsic curvature $\tilde{A}_{ij}=e^{-4\phi}(K_{ij}-\gamma_{ij}\,K/3)$, (d) the trace of the
extrinsic curvature $K$, and (e) the three auxiliary variables $\tilde{\Gamma}^i=-\partial_j
\tilde{\gamma}^{ij}$. These variables are evolved using the equations of motion (9)-(13) in
\cite{Etienne:2007jg}, along with the $1+$log time slicing for $\alpha$ and the
``Gamma--freezing" condition for $\beta^i$ cast in first order form (Eq.~(2)-(4)
in~\cite{Etienne:2007jg}). For numerical stability, we set the damping parameter $\eta$
appearing in the shift condition to $\eta=3.3/M$ for configurations with mass ratio $q=3:1$
and to $\eta=1.2/M$ for the configuration with mass ratio $q=5:1$ (see~Table~\ref{table:BHNS_ID}).
Here $M$  is the Arnowitt-Deser-Misner (ADM) mass of the system.

The spatial discretization is performed by using  fourth-order accurate, cell-centered,
finite-differencing stencils, except on shift advection terms, where fourth-order accurate
upwind stencils are used~\cite{Etienne:2007jg}. Outgoing wave-like boundary conditions are
applied to all the evolved variables. The time integration is performed via the method of
lines using a fourth-order accurate, Runge-Kutta integration scheme. Fifth order Kreiss-Oliger
dissipation~\cite{goddard06} has been also added in the BSSN evolution equations outside the
BH apparent horizon to reduce high-frequency numerical noise.
%
%
\begin{center}
  \begin{table}[th]
    \caption{Summary of the initial properties of the BHNS configurations. We list the mass ratio
      $q\equiv M_{\rm BH}/M_{\rm NS}$, where $M_{\rm BH}$ and $M_{\rm NS}$ are the masses
      of the BH and NS at infinite separation~(see~\cite{TBFS06} for details), the dimensionless
      BH spin parameter $\tilde{a}$,
      which is either aligned or anti-aligned with respect to the total angular momentum of the system,
      the dimensionless ADM mass $\bar{M}\equiv\kappa^{-1/2}\,M$ (here $k$ is the polytropic gas constant)
      and ADM angular momentum $J$ of the binary system, the orbital angular velocity $\Omega_0$, and
      a rough estimate of the innermost stable circular orbit (ISCO) separation $R_{\rm ISCO}$ computed
      via Eq.~(2.21) in~\cite{Bardeen72ApJ}. All the  NSs in the configurations have a nondimensional rest-mass
      ${\bar{M}_{\rm NS}}=0.15$. The label for each configuration includes successively:  a magnetic field
      configuration tag (Ali=aligned or Til=tilted), a tag identifying the binary mass ratio ($q=3$
      or $q=5$), and a tag identifying the spin direction (sp=aligned and sm-antialigned) and the
      magnitude of the BH spin. In all cases, the initial $M\,\Omega_0$ corresponds to an orbital
      separation of about $D_0\simeq 8.7M$.
      \label{table:BHNS_ID}}
    \begin{tabular}{ccccccc}
      \hline\hline
          Model                   & $q$ & $\tilde{a}$   & $\bar{M}$     &$J/M^2$ &  $M\,\Omega_0$ & $R_{\rm ISCO}/M_{\rm BH}$   \\  
          \hline
           Tilq3sp0.75       &  3  &  0.75         & 0.55          & 1.09     & 0.0328   & 3.2\\
           Aliq3sp0.5        &  3  &  0.5          & 0.55          & 0.96     & 0.0330   & 4.2\\          
           Aliq3sp0.0        &  3  &  0.0          & 0.55          & 0.70     & 0.0333   & 6.0 \\
           Aliq3sm0.5        &  3  & -0.5          & 0.55          & 0.44     & 0.0338   & 7.5 \\
           Aliq5sp0.0        &  5  &  0.0          & 0.83          & 0.52    & 0.0333    & 6.0 \\
          \hline\hline
    \end{tabular}
  \end{table}
\end{center}
%
\paragraph*{\bf MHD evolution:} For the matter and magnetic field, the Illinois code solves
the equations of ideal GRMHD in a conservative scheme via high-resolution shock capturing
methods. For that it adopts the conservative variables
\begin{eqnarray}
  \rho_* &\equiv &- \sqrt{\gamma}\, \rho_0\,n_{\mu}\,
  u^{\mu}\,,\,\,\,\,\,
  \tilde{\tau}\equiv\sqrt{\gamma}\, T_{\mu \nu}\,
  n^{\mu}\,n^{\nu} - \rho_*\,,
  \nonumber\\
  \tilde{S}_i &\equiv&-\sqrt{\gamma}\, T_{\mu \nu}\,
  n^{\mu}\,\gamma^{\nu}_i\,,
  \nonumber\label{eq:ConVar}
\end{eqnarray}
with $T_{\mu \nu}$ the stress-energy tensor for a magnetized plasma
defined as
\begin{equation*}
  T_{\mu \nu} = (\rho_0\,h+b^2)\,u_\mu\,u_\nu +
  \left( P + \frac{b^2}{2}
  \right)\,g_{\mu \nu} - b_{\mu}\,b_{\nu}\,,
  \label{eq:Tmunu}
\end{equation*}
where $\rho_0$ is the rest-mass density, $P$ is the pressure, $h=1+\epsilon+ P/\rho_0$ the specific
enthalpy, $\epsilon$  is the specific internal energy, $b^\mu=B^\mu_{(u)}/(4\,\pi)^{1/2}$  gives  the
magnetic field as measured by an observer co-moving with the fluid, $b^2=b^\mu\,b_\mu$ gives the magnetic
energy ($b$ is proportional to the magnitude of the magnetic field), and $u^\mu$
denotes the four-velocity of the fluid. We evolve the conservative
variables through  Eqs. (27)-(29) in~\cite{Etienne:2010ui}. To ensure the magnetic field remains
divergenceless during the evolution, we integrate the magnetic induction equation by introducing
a vector potential $\mathcal{A}^\mu$~(see Eqs. (19)-(20) in~\cite{Etienne:2010ui}). As noted
before~\cite{Giacomazzo:2012iv,Etienne:2010ui}, interpolations performed on the vector potential
at refinement boundaries on nested grids can induce spurious magnetic fields. To avoid that, we also
adopt the generalized Lorenz gauge~\cite{Farris:2012ux} with a damping parameter~$\xi\sim5.5/M$
for configurations with mass ratio $q=3:1$ and to $\xi=6.4/M$ for the configuration with mass ratio
$q=5:1$ (see~Table~\ref{table:BHNS_ID}). 
Finally, we adopt the  $\Gamma$--law EOS $P=(\Gamma-1) \rho_0\,\epsilon$, with $\Gamma=2$.
%
\subsection{Initial data}
\label{subsec:idata}
The quasiequilibrium BHNS configurations (see Table~\ref{table:BHNS_ID}) are constructed by
solving the GR constraint equations in the conformal thin-sandwich (CTS) decomposition, along
with the relativistic equations of hydrostatic equilibrium, imposing BH equilibrium boundary
conditions as in~\cite{CP04}. These CTS initial data correspond to BHNS binaries in a quasicircular
orbit with a separation chosen to be outside the tidal disruption radius~\cite{TBFS07b}. 

The initial data are calculated using the {\tt Lorene} spectral numerical libraries
\cite{2016ascl.soft08018G} employing dimensionless quantities as in~\cite{BSBook} where, for
example, the mass $M$ can be rescaled as $\bar{M}=k^{-1}\,M$, the spatial coordinates as 
$\bar{x}^i=k^{-1}\,x^i$, etc, where $k$ is the polytropic gas constant. The excised BH region is
populated with smooth junk data using the technique described in~\cite{EFLSB}. As in the previous
studies~\cite{Etienne:2007jg,Etienne:2011ea}, the initial data quantities are extrapolated from
the BH exterior into the interior using a $7^{th}$ order polynomial with a uniform stencil
spacing of $\Delta r\approx 0.3\,R_{\rm BH}$, with $R_{\rm BH}$ the radius of the apparent horizon.
A detailed description of our methods can be found in~\cite{TBFS07b,Etienne:2007jg}.

We assume that the initial NS can be modeled as an irrotational $\Gamma=2$ polytrope, and treat
BHs that are nonspinning ($\tilde{a}=0$), aligned ($\tilde{a} = 0.5$ and $0.75$)
and anti-aligned ($\tilde{a}=-0.5$) with respect to the total orbital angular momentum of the
system. The mass ratio considered here ranges from $q=3:1$ to  $q=5:1$ (see Table
\ref{table:BHNS_ID}). In all BHNS cases considered here the resulting NS has a compaction
of ${\cal C}={\mathcal M}_{\rm NS}/R_{\rm NS}=0.145$, where $\mathcal{M}_{\rm NS}$ and $R_{\rm NS}$ are
the ADM mass and  the  circumferential radius of the NS in isolation. For the adopted EOS
the maximum mass configuration has  $\mathcal{C}=0.215$. We rescale the  rest
mass of the star as $M_{\rm NS}=1.4M_\odot(k/189.96\rm km^2)^{1/2}$. For an isolated NS with
compaction ${\cal C}={\mathcal M}_{\rm NS}/R_{\rm NS} =0.145$, the ADM mass turns out to be
$\mathcal{M}_{\rm NS}=1.30 (M_{\rm NS}/1.4 M_\odot)M_\odot$, the isotropic radius is $R_{\rm iso}=
11.2(M_{\rm NS}/1.4M_\odot)\rm km$ and the Schwarzschild radius is $R_{\rm NS}=13.2(M_{\rm NS}/1.4M_\odot)\rm {km}$.
The maximum  rest-mass density of the NS is $\rho_{0,\rm max} =8.92\times 10^{14}\,(1.4M_\odot
/M_{\rm NS})^2\rm g/cm^{3}$. In all our BHNS configurations, the initial orbital angular velocity
$M\,\Omega_0$ corresponds to an orbital separation of about $D_0\approx 8.72M \sim 67.15
(M_{\rm NS}/1.4M_\odot)\rm km$ for configurations with mass ratio $q=3:1$, and $D_0\approx
8.68M\sim 101.35(M_{\rm NS}/1.4M_\odot)\rm km$ for mass ratio $q=5:1$. Note that these  BHNS
configurations have been used in~\cite{Etienne:2007jg,Etienne:2011ea}.

\begin{table*}
  \caption{Grid hierarchy for  models listed in Table~\ref{table:BHNS_ID}. Symmetry about the orbital plane
    (i.e. $z=0$) is imposed in all cases except in $\rm Tilq3sp0.75$ ($90^o$--tilted magnetic field)
    where we consider full 3D domain. The computational mesh consists of two sets of nested refinement boxes,
    one centered on the BH and the other on the NS. The finest box around the BH (NS) has a half length of
    $\sim 1.5\,R_{\rm BH}\,(1.2\,R_{\rm NS})$, where  $R_{\rm BH}\,(R_{\rm NS})$ is the initial radius of the
    BH (NS). The number of grid points covering the radius of the BH apparent horizon and the equatorial radius of
    NS is denoted by $N_{AH}$ and $N_{\rm NS}$, respectively. Note that the resolution used here matches that in
    Paper I, but it is higher than that in
    \cite{Etienne:2007jg,Etienne:2011ea} where the same cases were evolved.}
\begin{tabular}{cccccc}
  \hline \hline
  Model &  Grid Hierarchy (in units of $M$)$^{(a)}$  & Max. resolution & $N_{\rm AH}$ & $N_{\rm NS}$ \\
  \hline \hline
   Tilq3sp0.75  & (211.3, 93.0, 46.5, 23.2, 11.6, 5.8, 2.9, 1.45 [1.65], 0.76 [N/A]) & $M/60.6$ & 38 & 42 \\
   Aliq3sp0.5   & (253.6, 93.0, 46.5, 23.3, 11.6, 5.8, 2.9, 1.45 [1.65], 0.85 [N/A]) & $M/60.6$ & 35 & 42 \\
   Aliq3sp0.0   & (253.6, 93.0, 46.5, 23.3, 11.6, 5.8, 2.9, 1.45 [1.65], 0.96 [N/A]) & $M/60.6$ & 38 & 42 \\
   Aliq3sm0.5   & (253.6, 93.0, 46.5, 23.3, 11.6, 5.8, 2.9, 1.45 [1.65], 0.85 [N/A]) & $M/60.6$ & 35 & 42 \\
   Aliq5sp0.0   & (196.7, 98.3, 49.2, 24.6, 12.3, 4.4,2.2, 1.1)                      & $M/48.2$ & 41 & 48 \\
  \hline \hline
\end{tabular}
\begin{flushleft}
  $^{(a)}$  Half length of the refinement boxes centered on both the BH and the NS. When
  the side around the NS is different, we specify the NS half length in square brackets,
  or as $\rm [N/A]$ if there is no corresponding refinement box, i.e. if the NS is significantly larger
  than the BH.
\end{flushleft}
\label{table:grid}
\end{table*}

\begin{figure*}
  \centering
  \includegraphics[width=0.31\textwidth]{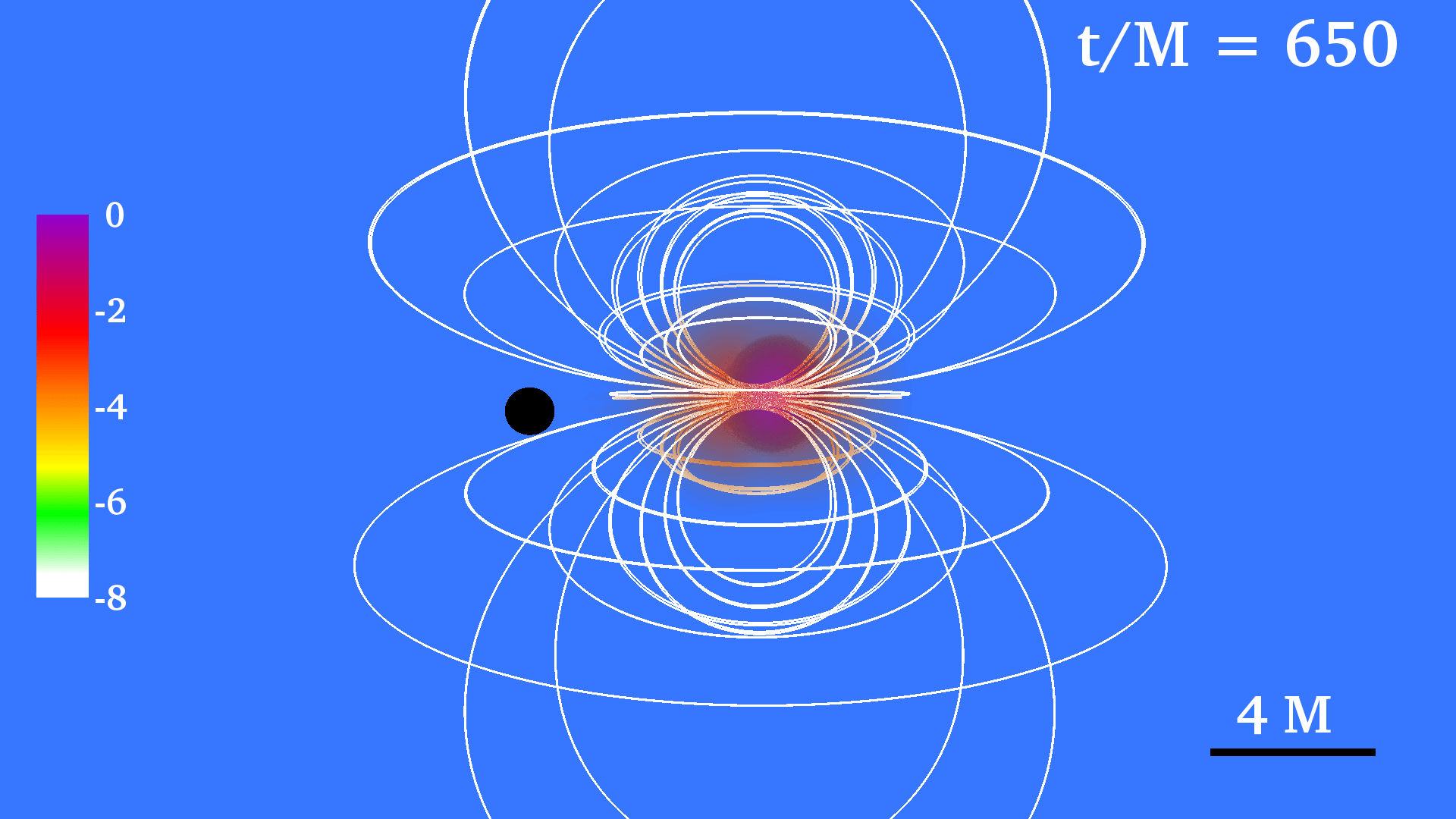}
  \includegraphics[width=0.31\textwidth]{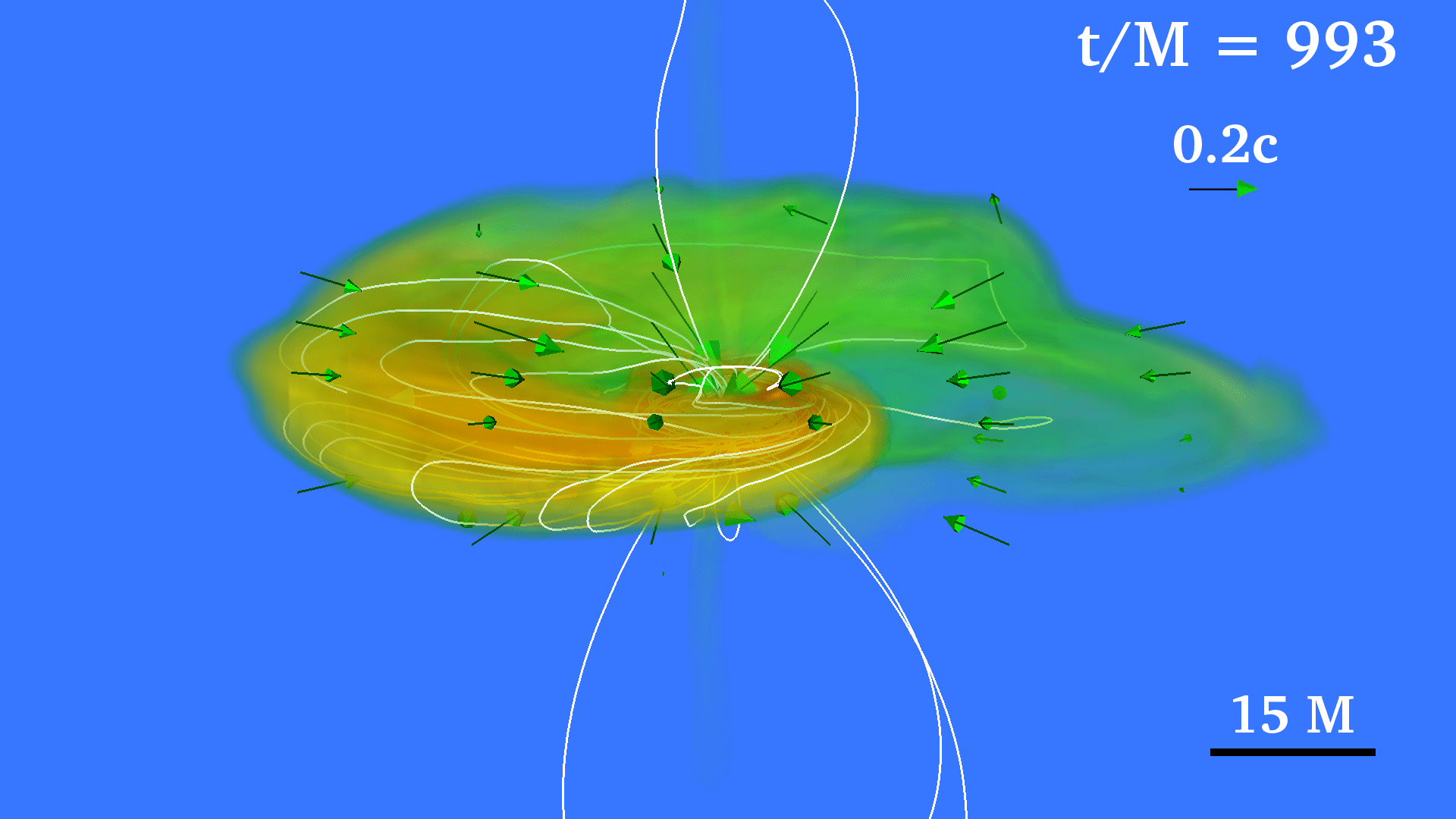}
  \includegraphics[width=0.31\textwidth]{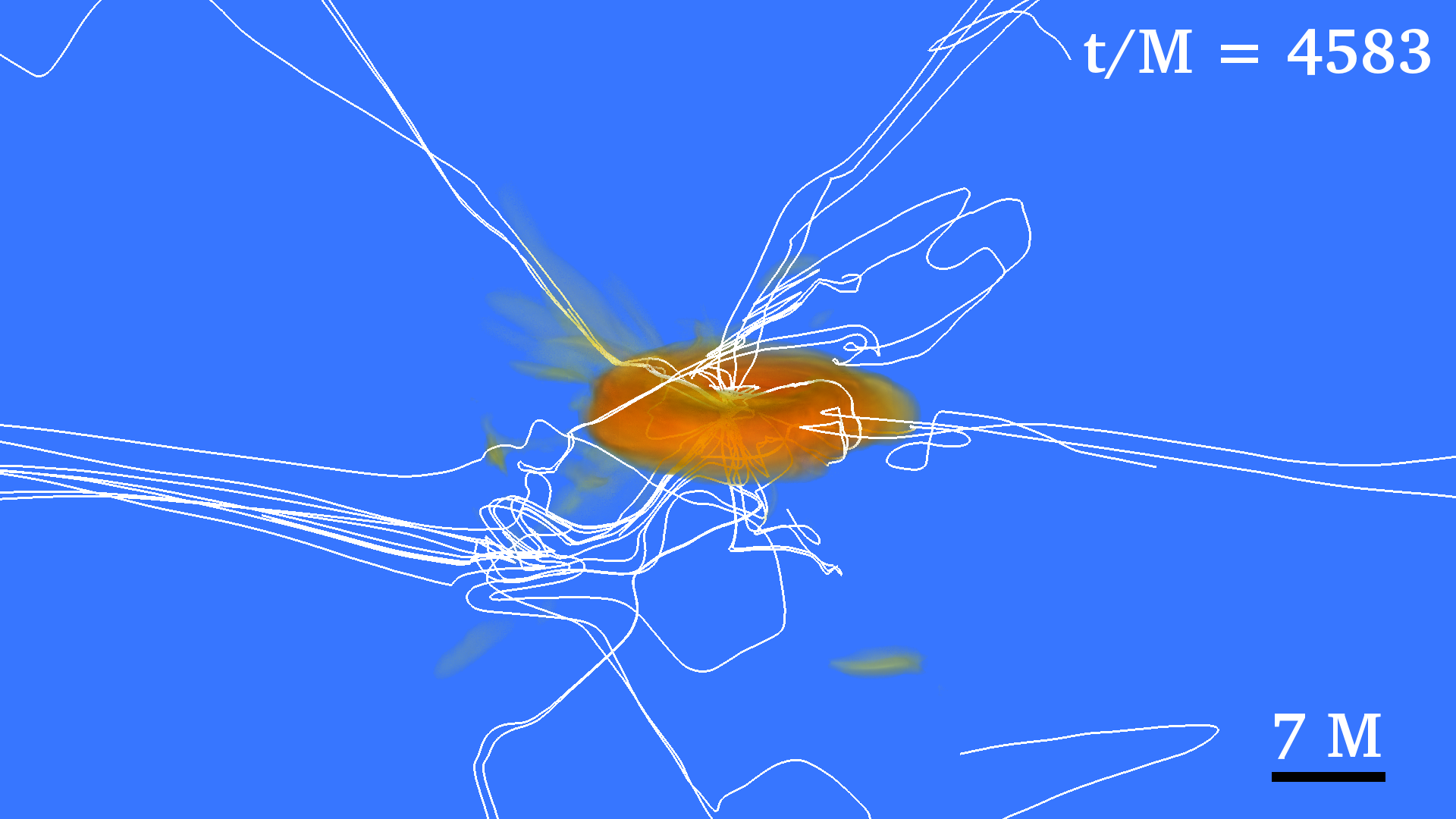}
\caption{Volume rendering of rest-mass density $\rho_0$ normalized to its initial NS
    maximum value  $\rho_0=8.92\times 10^{14}\,(1.4M_\odot/M_{\rm NS})^2\rm{g/cm}^{3}$
    (log scale) at selected times for case $\rm Aliq3sp0.0$ (see Table~\ref{table:BHNS_ID}).
    White lines denote the magnetic field while the arrows denote the fluid velocity. The
    BH apparent horizon is shown as a black sphere.  Here $M=2.5\times 10^{-2}(M_{\rm NS}/
    1.4M_\odot){\rm ms}=7.58(M_{\rm NS}/1.4M_\odot)\rm km$.
    \label{fig:BHNS_case_q_3_1_s075_tilted}}
\end{figure*}

Following~Paper~I, we evolve the configurations until they reach two orbits
prior to tidal disruption. At that point, the NS is endowed with a dynamically
unimportant, dipolar magnetic field generated by the vector potential
\cite{Paschalidis:2013jsa}
\begin{equation}
  A_\phi= \frac{\pi\,\varpi^2\,I_0\,r_0^2}{(r_0^2+r^2)^{3/2}}
  \left[1+\frac{15\,
    r_0^2\,(r_0^2+\varpi^2)}{8\,(r_0^2+r^2)^2}\right]\,, 
\label{eq:Aphi}
\end{equation} 
which approximately corresponds to a potential generated by an interior current loop.
Here $r_0$ is the current loop radius, $I_0$ is the current, $r^2=\varpi^2+z^2$,
with $\varpi^2 =(x-x_{\rm NS})^2 +(y-y_{\rm NS})^2$, and $(x_{\rm NS},y_{\rm NS})$ is the position
of the center of mass of the NS. As is displayed in~Table~\ref{table:BHNS_ID}, we
consider configurations in which the dipole magnetic moment is either aligned (see
left top panel in Fig.~\ref{fig:BHNS_case_q_3_1_s05}) or tilted by~$90^o$
(see left panel in Fig.~\ref{fig:BHNS_case_q_3_1_s075_tilted})~with respect
to the total orbital angular momentum of the system. 
%

For comparison purposes, we choose the current $I_0$ and radius of the loop $r_0$ such that
the magnetic pressure is $5\%$ of the gas pressure at the center of the NS as in~Paper~I.
The resulting magnetic field strength is $B_{\rm pole}\simeq 6.7\times 10^{15}(1.4M_\odot/
M_{\rm NS})$G on the surface of the star. Notice that although the resulting magnetic field
is large, it is still dynamically unimportant and, as it was shown in Paper~I, does not
affect  the tidal disruption or the merger phases. We expect therefore that the final outcome
of the post-merger phase should be approximately independent of the initial magnetic field
strength; the amplification of the magnetic field  following  disruption  is mainly due
to magnetic winding and the MRI~\cite{kskstw15}.

\begin{figure}
  \centering
  \includegraphics[width=0.48\textwidth]{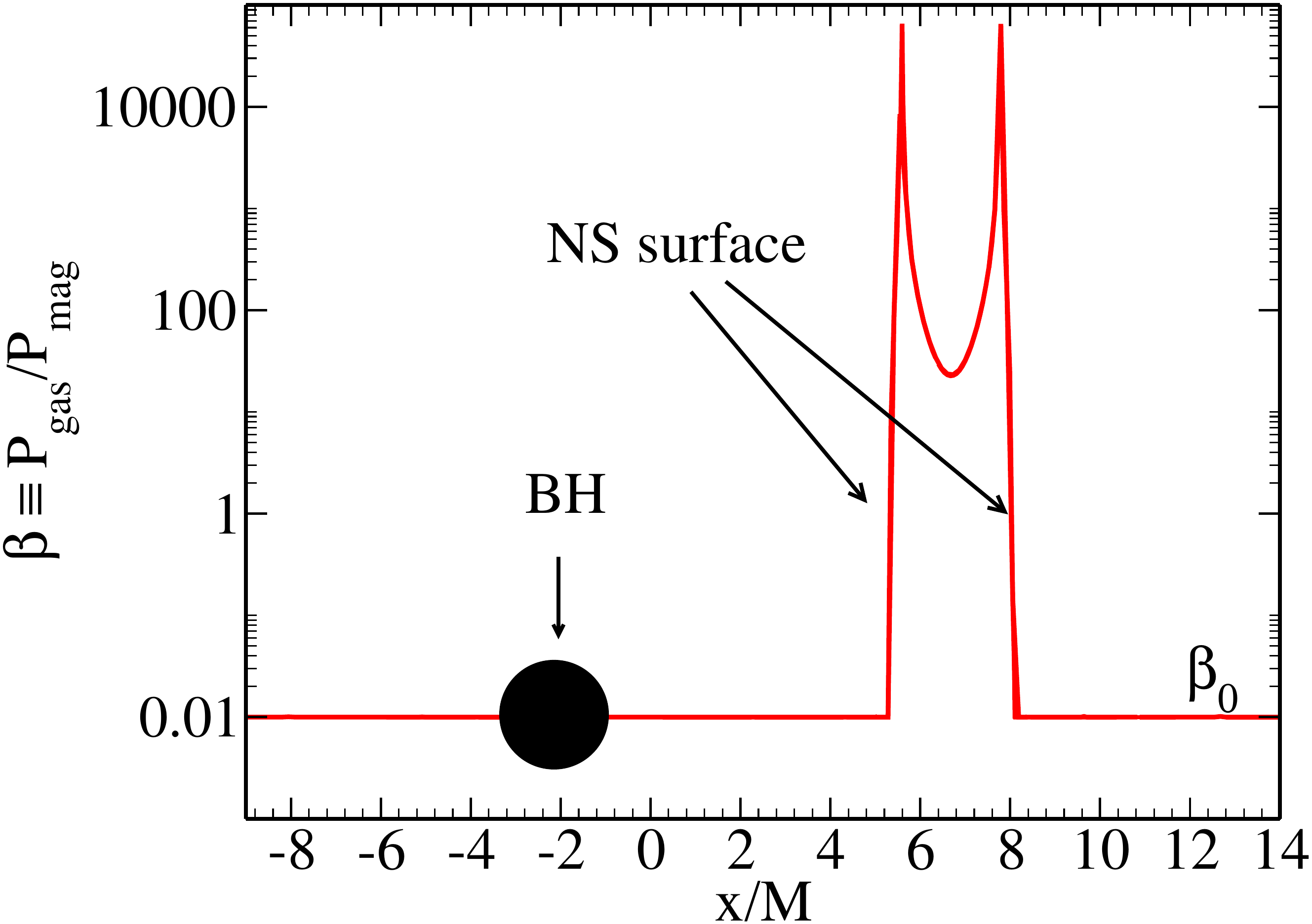}
  \caption{Gas-to-magnetic pressure ratio  $\beta\equiv P_{\rm gas}/P_{\rm mag}$
    along the x-direction at the time $t=t_B$ the dipole-like magnetic field generated
    by the vector potential $A_\phi$ in Eq.~\ref{eq:Aphi} is seeded in the star
    (see Table~\ref{table:BHNS_ID}).
    \label{fig:plasma_ext}}
\end{figure}

To reliably evolve  the  exterior magnetic field with the Illinois GRMHD code,
and at the same time mimic the magnetic-pressure dominant environment that likely
characterizes the force-free, pulsar-like exterior magnetosphere at the time the magnetic
field is seeded in the NS ($t=t_B$), a low and variable density  is enforced initially in regions
where magnetic field stresses dominate over the fluid pressure gradient. This procedure
is typically done in ideal MHD codes to evolve exterior magnetic fields~(see e.g.
\cite{Font:2007zz}).  This ``atmosphere'' is constructed such that the
exterior gas-to-magnetic-pressure ratio (the plasma parameter $\beta$) equals a target
value~$\beta_0\ll 1$ everywhere~(see Fig.~\ref{fig:plasma_ext}).
This choice allows us to automatically define the NS surface as the region where the
interior plasma parameter $\beta$ equals $\beta_{0}$ for the first time in moving outward
from the center, or equivalently
\begin{equation}
  \rho_0^{\rm surf} =\left({\frac{\beta_0\,b^2}{2\,\kappa}}\right)^{1/\Gamma}
  \ll \rho_{0,c}\,,
\end{equation}
with $\Gamma=2$, and $\rho_{0,c}$ the initial NS central density. In the stellar exterior
we reset the rest-mass density to $\rho_0=\rho_0^{\rm surf}$. The profile for $\beta$ both
inside the star, where the field is weak, and outside is plotted in Fig.
\ref{fig:plasma_ext}. The density outside at $t=t_B$ is set to
\begin{equation}
  \rho_0^{\rm atm}=\left({\frac{\beta_0\,b^2}{2\,\kappa}}\right)^{1/2}\,,
  \label{eq:extNS}
\end{equation}
so that as the magnetic field strength falls from the NS surface as $1/r^3$,
the above prescription forces $\rho_0^{\rm atm}$ to fall as $1/r^3$ as well.

%

In Paper I we showed that  different exterior conditions ranging from moderate
to complete magnetic field pressure dominance ($\beta_{0}=0.1,\,0.05,\,0.01$) do not affect
the final outcome of the BHNS mergers; a larger $\beta_0$ affects the inertia of the matter
in the atmosphere resulting in a delayed jet launching. We set $\beta_0=0.01$ which provides
the best approximation to a force-free environment that our code can handle reliably. This
choice of~$\beta_{0}$  increases total rest-mass of the system in less than $1\%$. 

We assume that the pulsar-like magnetosphere comoves with the NS, for which we set
the exterior plasma three-velocity to
\begin{equation}
  v^i=
  \begin{cases}
    v^i_{\rm CM}\,,& \text{if}\,\,\, \varpi\le 3\,R_{\rm NS}\,,\\
    v^i_{\rm CM}\,(3\,R_{\rm NS}/\varpi)^4-&\\
    \beta^i\,(1-(3\,R_{\rm NS}/\varpi)^4)\,,& \text{if}\,\,\, \varpi> 3\,R_{\rm NS}\,,
    \end{cases}
  \end{equation}
where $v^i_{\rm CM}$ is the three-velocity of the NS centroid.  This condition implies that the
variable atmosphere is stationary with respect to Eulerian observers.

For the subsequent evolution, we integrate  the ideal GRMHD equations everywhere, imposing a
density floor in regions where where $\rho_0^{\rm atm}< 10^{-10} \rho_0^{max}$, where
$\rho_0^{max}$ is  the initial maximum density of the NS.
%
\subsection{Grid structure}
\label{subsec:grid}
%
The grid hierarchy used in our simulations is summarized in Table~\ref{table:grid}. It consists
of two sets of mesh nested refinement boxes centered on both the BH and the NS. We use 9 nested
boxes centered on the BH and 8 boxes centered on the NS in configurations with mass ratio $q=3:1$,
and 8 nested boxes centered on the BH and on the NS in the configuration with  mass ratio $q=5:1$.
The finest box has a half length of $\sim 1.5\,R_{\rm BH}$ around the BH and $\sim 1.2\,R_{\rm NS}$
around the NS. These choices resolve the initial apparent horizon equatorial radius by $\gtrsim 70$
grid points, and the initial NS equatorial radius by $\gtrsim 84$ grid points.  We impose reflection
symmetry across the orbital plane ($z=0$) for all configurations for which the magnetic dipole
moment is aligned with the orbital angular momentum of the system, and consider the full 3D domain
for the $90^o$--tilted magnetic field (see Table~\ref{table:BHNS_ID}).  Note that the resolution
employed here matches the one used in Paper I, and it is higher than that previously employed
in~\cite{Etienne:2007jg,Etienne:2011ea} where same cases were evolved.
%
%
\begin{table*}[]
  \begin{center}
    \caption{Summary of main results. Here $\tilde{a}$ is the dimensionless remnant BH spin parameter,
      $\Delta E_{\rm GW}$ and $\Delta J_{\rm GW}$ are the total energy and angular momentum carried away by GWs,
      respectively. The kick velocity due to recoil is denoted by $v_{\rm kick}$ in $\rm km/s$, $b^2/
      (2\rho_0)_{\rm ave}$ is the space-averaged value of the magnetic-to-rest-mass-density ratio (force-free
      parameter) over all the grid points inside a cubical region of length $2R_{\rm BH}$ above the BH pole (see
      Fig.~\ref{fig:b2_2rho0}), $B_{\rm rms}$ denotes the rms value of the magnetic field above the BH poles
      in units of $(1.4M_\odot/M_{\rm NS})$G, $\alpha_{SS}$ is the Shakura--Sunyaev viscosity parameter, $M_{\rm disk}$
      is the rest-mass of the accretion disk remnant, $\dot{M}$ is the rest-mass accretion rate computed via Eq.
      (A11)~in~\cite{Farris:2009mt}, $\tau_{\rm disk}\sim M_{\rm disk}/\dot{M}$ is the disk lifetime (lifetime
      of the jet, if any) in units of $(M_{\rm NS}/1.4M_\odot)$s, and $L_{\rm jet}$ is the Poynting luminosity in units
      of $\rm erg/s$ driven by the incipient jet, time-averaged over the last $500M\sim 12.5(M_{\rm NS}/1.4M_\odot)\rm ms$
      of the  evolution. A dash  denotes ``no information available''.
      \label{table:results_allBHs}}
    \begin{tabular}{cccccccccccc}
      \hline\hline
          {Model} &  $\tilde{a}$ &$\Delta E_{\rm GW}/M_{\rm ADM}$ &$\Delta J_{\rm GW}/J_{\rm ADM}$& $v_{\rm kick}$ &$b^2/(2\rho_0)_{\rm ave}$ & $B_{\rm rms}$ &
          $\alpha_{SS}$&$M_{\rm disk}/{M_{\rm NS}}$ &
          $\dot{M} (M_\odot/s)$ & $\tau_{\rm disk}$
          & $L_{\rm jet}$\\
          \hline
          Aliq3sp0.75$^{(a)}$   & 0.85  & $0.97\%$& $14.25\%$ & 54.20 & $\gtrsim 100$     & $\gtrsim 10^{15.0}$ & $0.01-0.03$  & $10.0\%$   &  0.25  & $0.5$  & $10^{51.2}$ \\          
          \hline 
          Tilq3sp0.75   & 0.85  & $1.0\%$ & $14.33\%$ & 54.34 & 0.26      & $10^{14.1}$  & $0.01-0.013$     & $11.29\%$  &  0.29  & $0.54$   & $-$           \\
          Aliq3sp0.5    & 0.76  & $0.96\%$& $14.95\%$ & 65.32 & 113.7     & $10^{15.5}$  & $0.012-0.031$    & $6.15\%$   &  0.12  & $0.71$  & $10^{51.6}$   \\
          Aliq3sp0.0    & 0.54  & $1.0\%$ & $18.38\%$ & 45.20 & 3.25      & $10^{14.6}$  & $0.013-0.022$    & $2.33\%$   &  0.09  & $0.36$  & $-$           \\
          Aliq3sm0.5    & 0.33  & $0.99\%$& $24.96\%$ & 56.65 & $10^{-3}$ & $10^{13.3}$  & $-$              & $0.24\%$   &  0.03  & $0.11$  & $-$           \\
          Aliq5sp0.0    & 0.41  & $0.91\%$& $19.63\%$ & 69.96 & $10^{-3}$ & $10^{12.3}$  & $-$              & $0.34\%$   &  0.04  & $0.12$  & $-$           \\
          \hline\hline
    \end{tabular}
  \end{center}
  \begin{flushleft}
    $^{(a)}$ BHNS configuration reported in~paper~I for $\beta_0=0.01$.
  \end{flushleft}
\end{table*} 
%
\subsection{Diagnostic quantities}
\label{subsec:diagnostics}
During the numerical integration we adopt  a number of  diagnostics to analyze
and verify the reliability of our magnetized BHNS mergers. We monitor the $L_2$
normalized Hamiltonian and momentum constraints  computed via  Eqs.~(40)-(41)
in~\cite{Etienne:2007jg}. In all cases listed in Table~\ref{table:BHNS_ID}, we
find that the constraint violations peak at $\lesssim 2.7\%$ during the merger,
as expected. During inspiral and post-merger phases, the violations are smaller
than $\lesssim 1\%$, and stay roughly constant until the end of the evolution.
The BH apparent horizon  is located  and  monitored through  the
{\tt AHFinderDirect} thorn~\cite{ahfinderdirect}. We estimate the BH mass
$M_{\text{\rm BH}}$ and the BH dimensionless spin parameter
$\tilde{a}$ via~Eqs.~(5.2)-(5.3) in~\cite{Alcubierre:2004hr}. We monitor the
conservation of both the total mass $M_{\rm int}$ and the total angular momentum
$J_{\rm int}$ interior to a large radius $r$,  which coincide with the ADM mass and
ADM angular momentum of the system at $r=\infty$, via Eqs. (19)-(22) in~\cite{Etienne:2011ea}.
To measure  the flux of energy and angular momentum  carried away  by GWs, we
use a  modified version of the {\tt Psikadelia} thorn that computes the  Weyl
scalar $\Psi_4$, which is decomposed into $s=-2$ spin-weighted spherical
harmonics~\cite{Ruiz:2007yx} at different radii between $r_{\rm min}\approx 22M\sim
166(M_{\rm NS}/1.4M_\odot)\rm km$ and $r_{\rm max}\approx 130M\sim 985 (M_{\rm NS}/
1.4M_\odot)\rm km$ for cases with mass ratio $q=3:1$,  and $r_{\rm min}\approx 22M
\sim 252(M_{\rm NS}/1.4M_\odot)$km, and $r_{\rm max}\approx 130M\sim 1490 (M_{\rm NS}/
1.4M_\odot)\rm km$ for the mass ratio $q=5:1$. We find that 
$\sim 1.0\%$  of the total energy of our BHNS models is radiated away during the
evolution in form of gravitational radiation, while between $\sim 14\%$ and
$\sim 25\%$ of the angular momentum is radiated (see~Table~\ref{table:results_allBHs}).
Taking  into account the GW radiation losses, we also find that, in all configurations
considered here, the violation of the conservation of~$M_{\rm int}$~is $\lesssim 1\%$
along the whole evolution, while the violation of the conservation of~$J_{\rm int}$
is $\sim 1\%$ in cases $\rm Aliq3sm0.5$ and~$\rm Aliq5sp0.0$ (see Table~\ref{table:BHNS_ID}),
and $\lesssim 4\%$ in the remaining cases.

In addition, we monitor the conservation of the rest-mass $M_{\rm NS}=\int \rho_* d^3x$, where
$\rho_*\equiv \sqrt{\gamma}\rho_0\,n_\mu\,u^\mu$, as well as the magnetic energy growth
outside the BH apparent horizon through
\begin{equation}
  \mathcal{M}=\int u^\mu u^\nu T^{(EM)}_{\mu\nu}\,dV\,,
  \label{eq:EM_energy}
\end{equation}
as measured by a comoving observer~\cite{Etienne:2011ea}, where $dV=e^{6\phi}\,d^3x$
is the proper volume element on the spatial slice. Here $T^{(EM)}_{\mu\nu}$
is the electromagnetic energy-momentum tensor. The rest-mass accretion rate is computed
via mass fluxes across the apparent horizon as 
\begin{equation}
  \dot{M}=-\int_{\rm AH} \alpha\,\sqrt{\gamma}\,\rho_0\,u^\mu
  \partial_\mu f\,J\,d\theta\,d\phi\,,
\label{eq:Mdot}
\end{equation}
where
\begin{eqnarray}
  f&=&\sqrt{(x-x_h(t))^2 + (y-y_h(t))^2 + (z-z_h(t))^2}\nonumber\\
  &-&R(t,\theta,\phi)\,,
\end{eqnarray}
is a scalar function such that $f=0$ on the spatial hypersurface
corresponding to the world tube of the BH apparent horizon. Here
$J=\partial(f,\theta,\phi)/\partial(x,y,z)$ is the Jacobian,
$(x_h,y_h,z_h)$ is the position of the BH centroid, and $R(t,\theta,\phi)$
represents the coordinate distance from the BH centroid to the apparent horizon
along the~$(\theta,\phi)$ direction. For details see Appendix~A~in~\cite{Farris:2009mt}.

To probe MHD turbulence  in our systems, we compute the effective
Shakura--Sunyaev $\alpha_{\rm SS}$ parameter~\cite{Shakura73} associated  with the effective
viscosity due to magnetic stresses through $\alpha_{\rm SS}\sim T^{EM}_{\hat{r}\hat{\phi}}/P$
(see Eq. 26 in~\cite{FASTEST_GROWING_MRI_WAVELENGTH}).  We also  verify that the MRI can be
captured in the post-merger phase of our simulations by computing the  quality factor
$Q_{\rm MRI}\equiv\lambda_{\rm MRI}/dx$, which  measures the number of grid points per fastest
growing MRI mode. Here $\lambda_{\rm MRI}$ is  the fastest-growing MRI wavelength defined
as~\cite{UIUC_PAPER2}
\begin{align}
\lambda_{\rm MRI} \approx 2\,\pi\,\frac{\sqrt{|b_{P}b^{P}|/(b^2+\rho_0\,h)}}{|\Omega(r,\theta)|},
\label{eq:lambda_MRI}
\end{align}
where $|b^{P}| \equiv \sqrt{b^2-b_\mu\,(e_{\hat\phi})^\mu|^2}$, and
$(e_{\hat\phi})^\mu$ is the orthonormal vector carried by an observer comoving
with the fluid, $\Omega(r,\theta)$ is the angular velocity of the disk remnant,
and $dx$ is the local grid spacing. Typically to capture MRI requires $Q_{\rm MRI}\gtrsim 10$
(see e.g.~\cite{Sano:2003bf,Shiokawa:2011ih}). Finally,  we compute the outgoing EM Poynting
luminosity
\begin{equation}
  L=-\int T^{r(EM)}_t\,\sqrt{-g}\,d\mathcal{S}\,,
  \label{eq:Lem}
\end{equation}
across spherical surfaces of coordinate radii between $R_{\rm ext}=46 M\simeq 350(M_{\rm NS}/
1.4M_\odot)$km and $190 M\simeq 1440(M_{\rm NS}/1.4M_\odot)$km. 
%
%
\section{Results}
\label{sec:results}
As all our initial BHNS binaries are in a quasicircular orbit with an initial
coordinate separation outside the tidal disruption distance, their evolution 
can be roughly characterized by three stages: late inspiral, tidal disruption-and-merger,
and post-merger. During the late inspiral, the orbital separation  decreases as energy
and angular momentum are carried off by gravitational radiation. Once the NS is disrupted 
a rapid redistribution of the angular momentum in the external layers of the star pushes
matter out of the innermost stable circular orbit (ISCO) causing long tidal tails
(see right top and left middle panels in~Fig.~\ref{fig:BHNS_case_q_3_1_s05}). Depending on
the specific angular momentum of the matter in the tidal tail, it can be accreted, it
can wrap around the BH to form the accretion disk (see left middle panel in
Fig.~\ref{fig:BHNS_case_q_3_1_s05}), or it can be dumped in the atmosphere as escaping 
or fall-back debris.

The fluid motion in the new-born disk drags the frozen-in magnetic field lines
into a predominantly toroidal configuration. However, the presence of an external
magnetic field in the initial NS that connects matter in the star with footpoints at
the poles of the BH establishes a poloidal field component that persists throughout
the disk and amplifies following tidal disruption (see second row in
Fig.~\ref{fig:BHNS_case_q_3_1_s05} and central panel in
Fig.~\ref{fig:BHNS_case_q_3_1_s075_tilted}). Depending on the poloidal
magnetic field, the fall-back debris, and the rest-mass of the disk, these instabilities
may induce high magnetic pressure gradients above the BH poles that eventually can launch
an outflow. In paper I, we showed for the first time that BHNS remnants with a strong
poloidal magnetic field component can launch a collimated, mildly relativistic
outflow---an incipient jet-- and hence be the progenitors of sGRBs. In the following
section, we summarize the dynamics of our new  BHNS configurations that differ in BH spin,
mass ratio,  and magnetic field configuration~(see~Table~\ref{table:BHNS_ID}).
Table~\ref{table:results_allBHs} highlights the key parameters at the termination of
our simulations.
%
\subsection{Effect of black hole spin}
\label{subsec:Ef_spin}
%
\begin{figure*}
  \centering
  \includegraphics[width=0.33\textwidth]{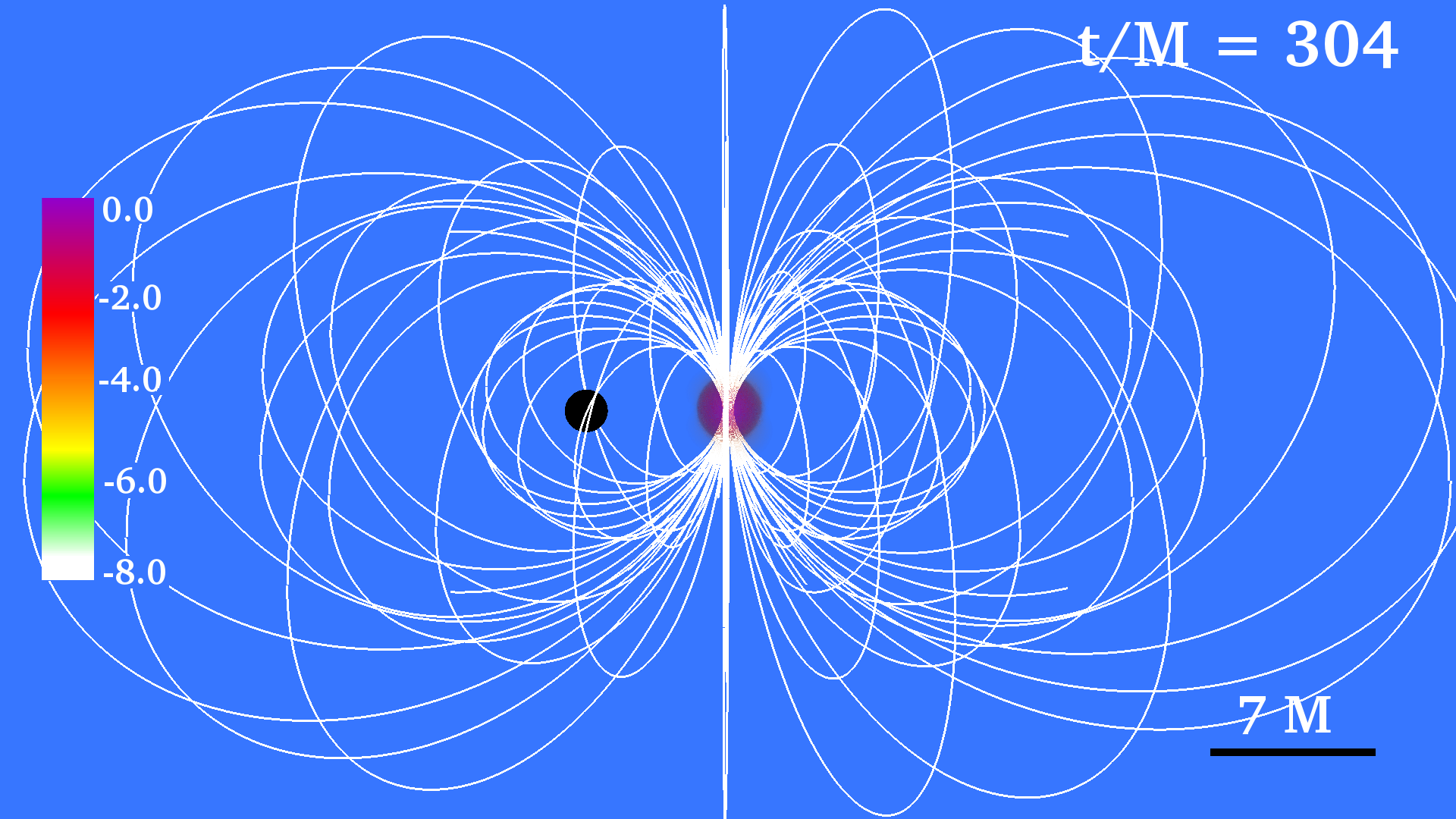}
  \includegraphics[width=0.33\textwidth]{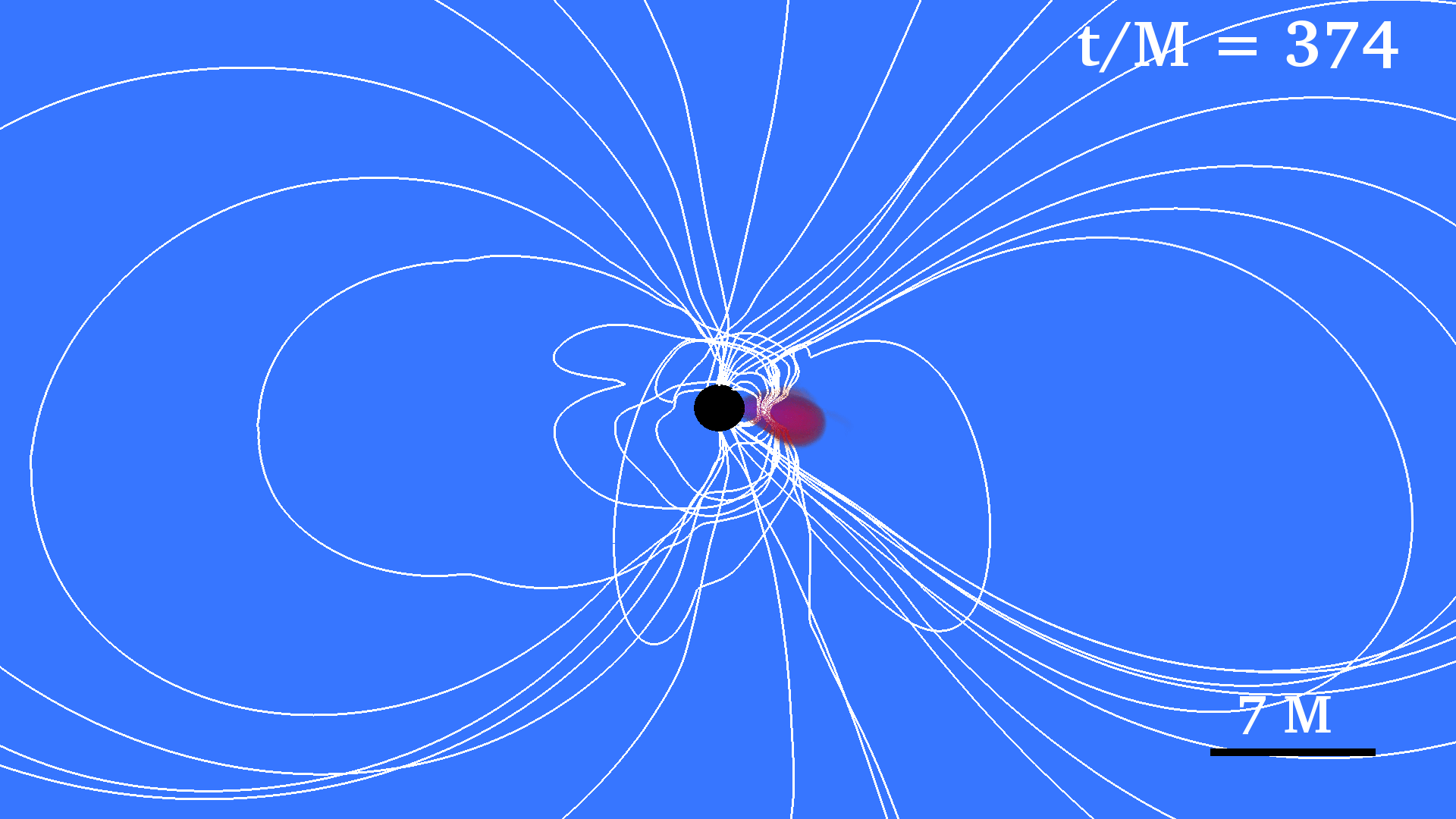}
  \includegraphics[width=0.33\textwidth]{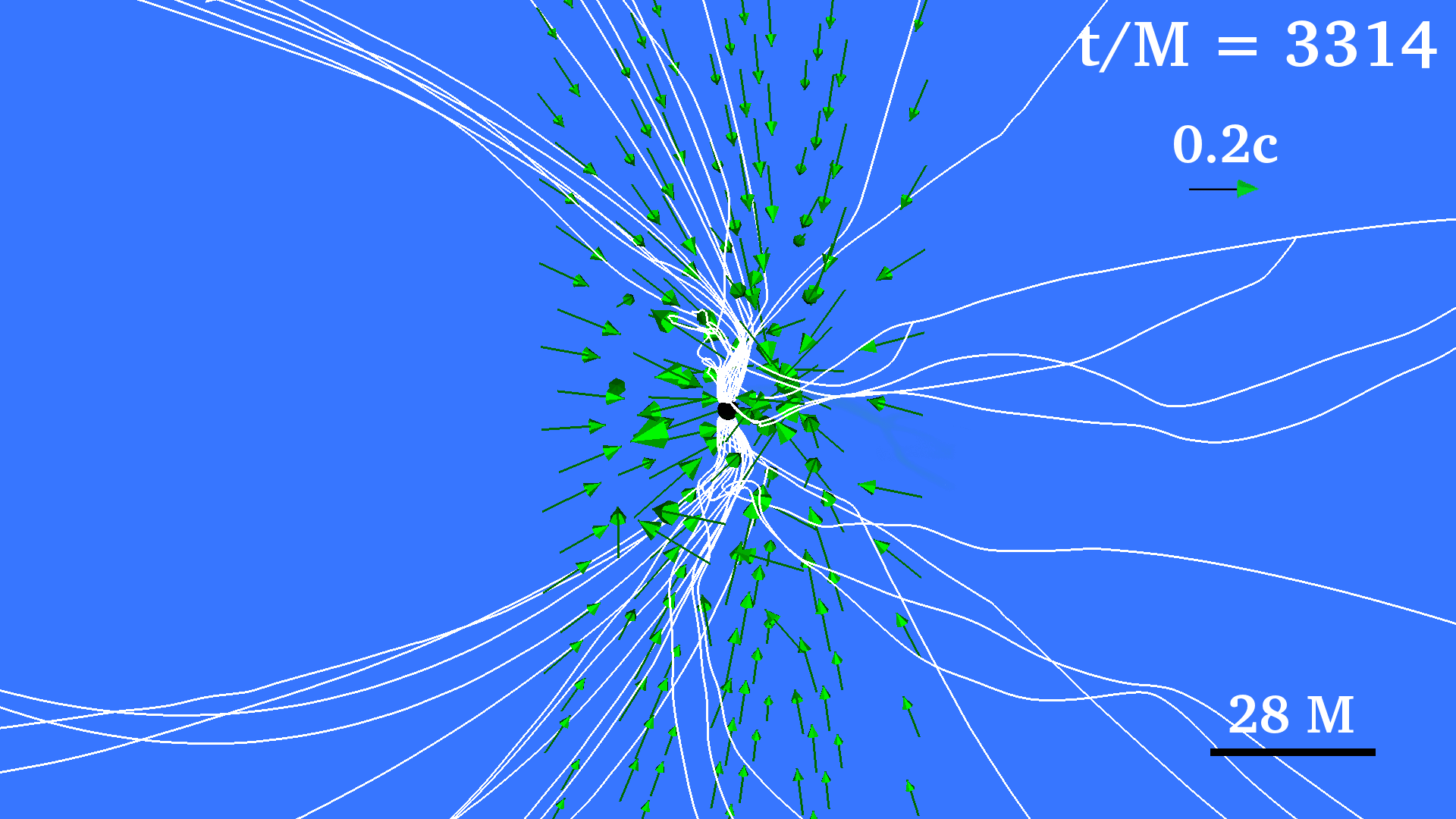}
  \includegraphics[width=0.33\textwidth]{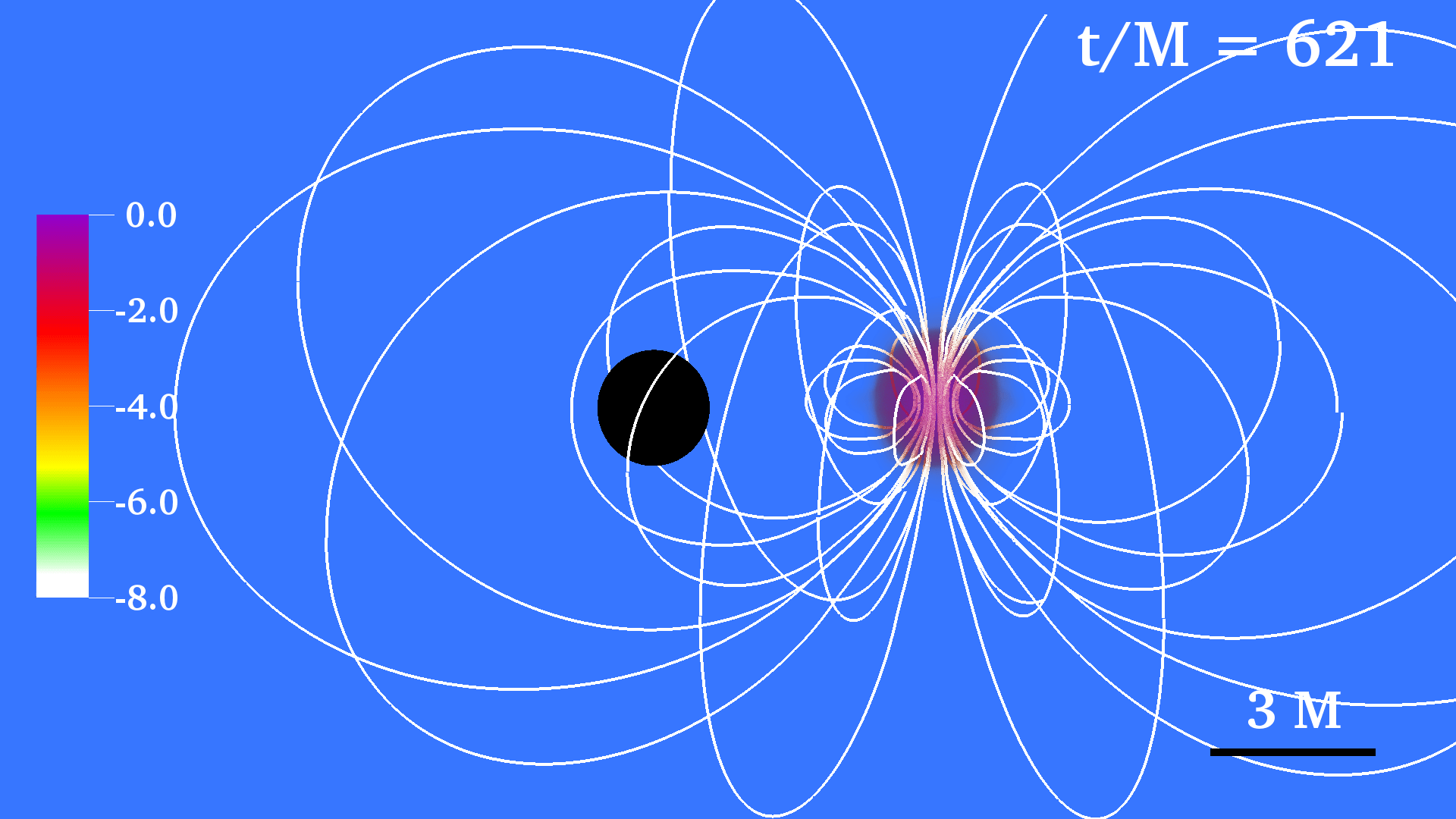}
  \includegraphics[width=0.33\textwidth]{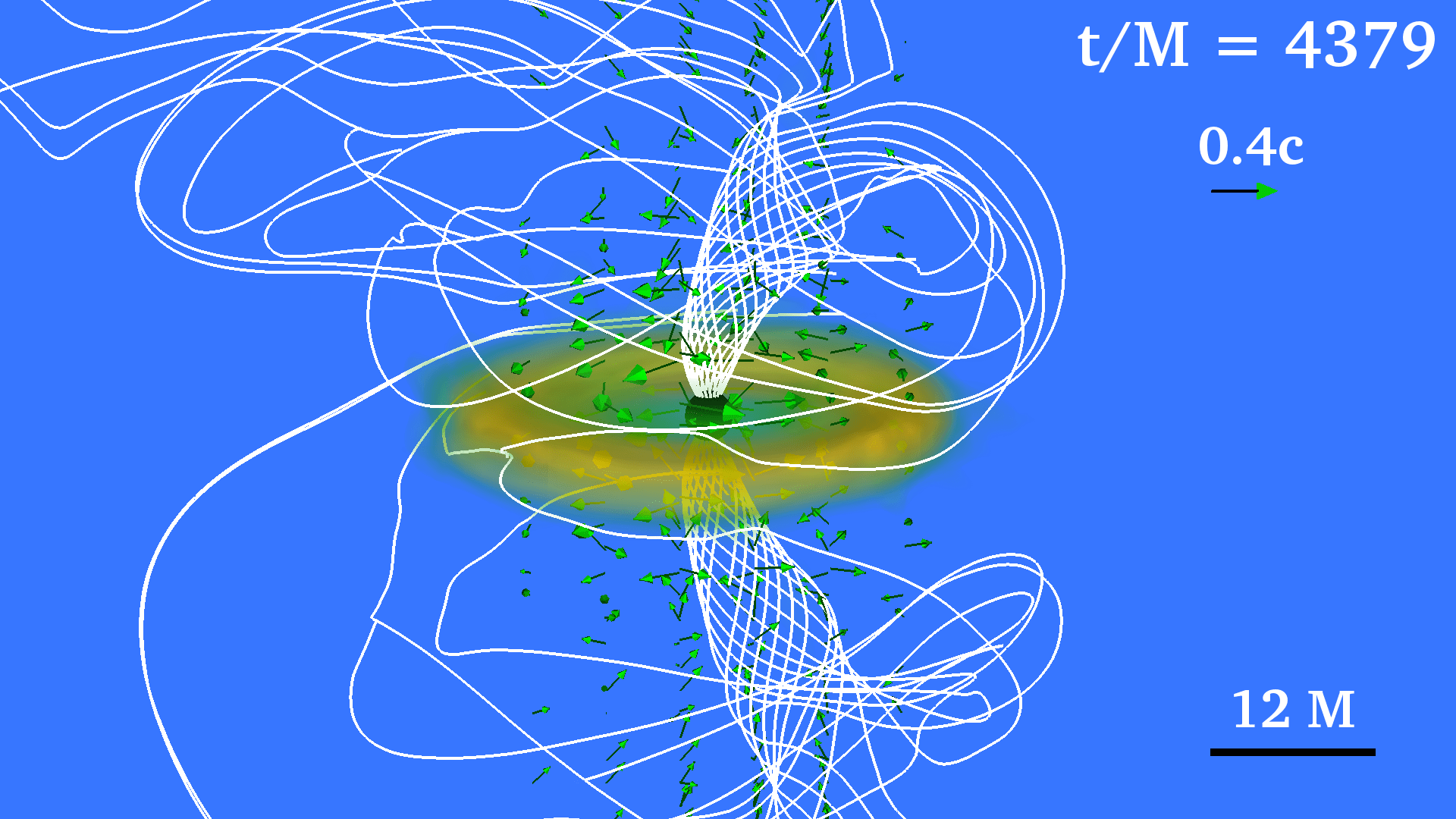}
  \includegraphics[width=0.33\textwidth]{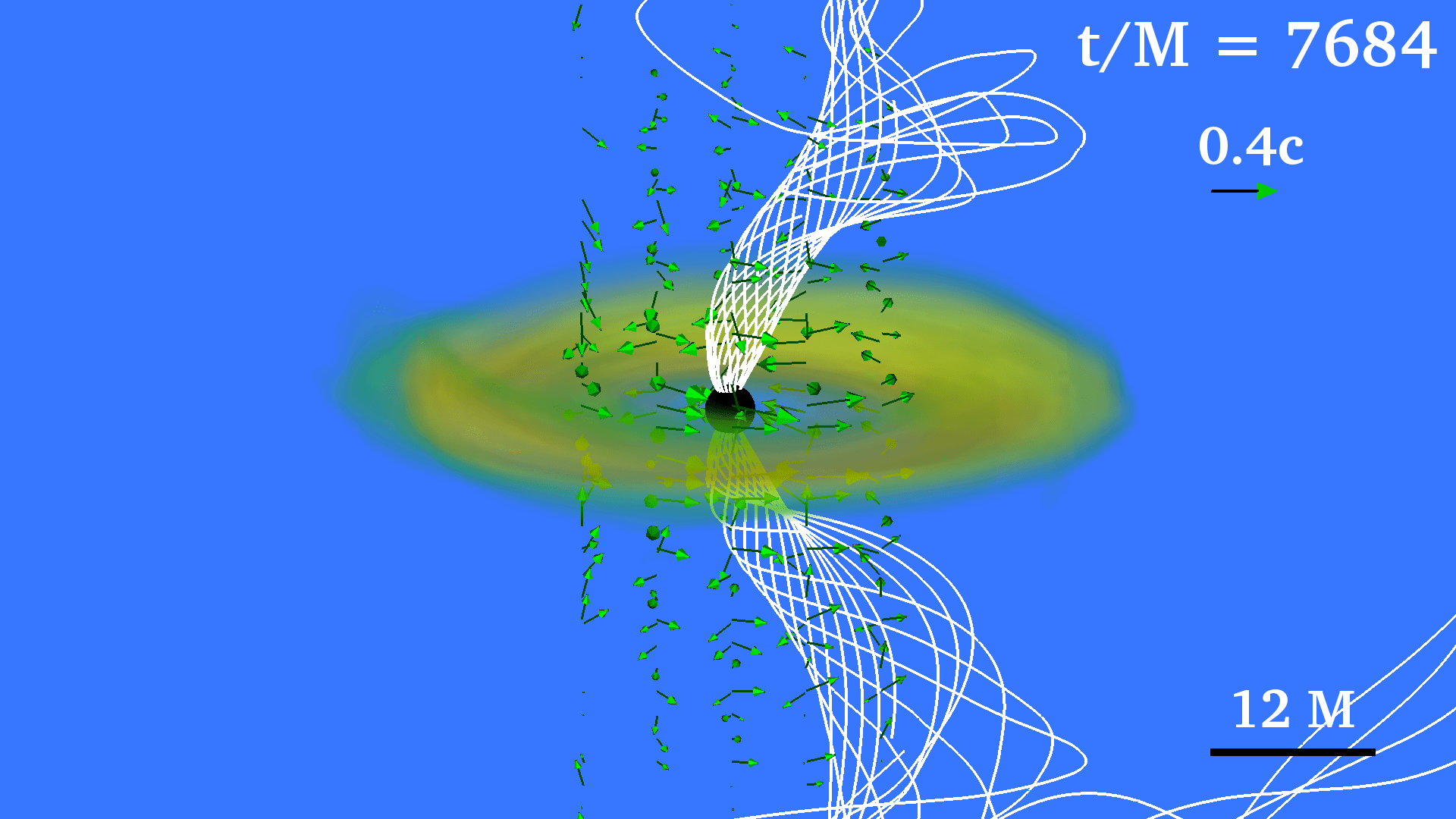}
  \caption{Volume rendering of rest-mass density $\rho_0$ normalized to its initial NS
    maximum value  $\rho_0=8.92\times 10^{14}\,(1.4M_\odot/M_{\rm NS})^2\rm{g/cm}^{3}$
    (log scale)  for cases Aliq3sm0.5 (top row) and Aliq3sp0.0 
    (bottom row) at selected times. White lines denote the magnetic field while the arrows
    denote the fluid velocity. The BH apparent horizon is shown as a black sphere.  Here $M=2.5\times
    10^{-2}(M_{\rm NS}/1.4M_\odot)\rm ms$= $7.58(M_{\rm NS}/1.4M_\odot)\rm
    km$.~\label{fig:BHNS_case_s0_sm05}}
    \end{figure*}
To disentangle the effects of the initial BH spin $\tilde{a}$ on jet launching
from the effects of the  mass ratio and the magnetic field geometry,  we next consider
only configurations with mass ratio $q=3:1$, aligned magnetic field and BH spin
$\tilde{a}=-0.5,\,0.0,\,0.5$. For comparison, we also summarize the results  of the
configuration reported in Paper I that corresponds to a similar configuration but with
a BH spin $\tilde{a}=0.75$.

Figs.~\ref{fig:BHNS_case_q_3_1_s05} and \ref{fig:BHNS_case_s0_sm05} (see also Fig.~1 in
Paper I) display snapshots of the evolution of the rest-mass density along with the magnetic
field lines starting from magnetic field insertion at $t=t_B$, followed by the disruption
of the star and the formation of the accretion disk.  The bulk of the star is accreted into
the BH, and the disk + BH remnant eventually settles down as does the outflow, when it occurs.
%
\begin{figure}
  \centering
  \includegraphics[width=0.49\textwidth]{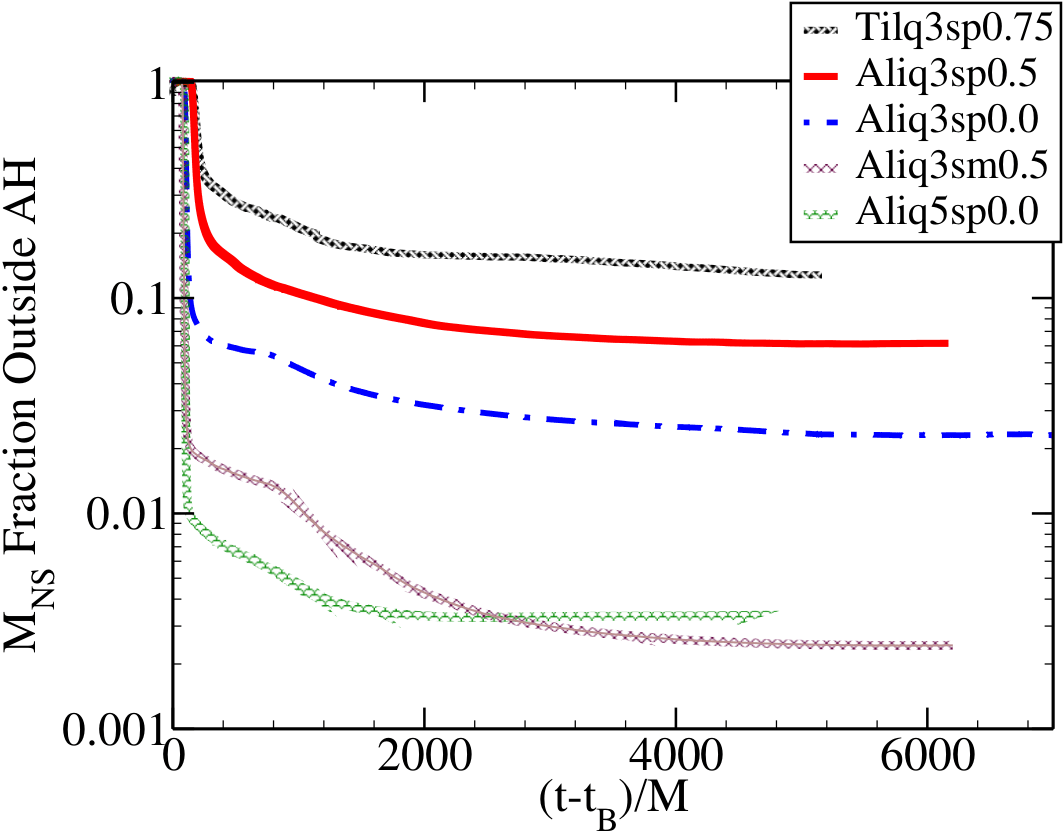}
  \caption{Rest-mass $M_{\rm NS}$ of NS matter outside the BH versus time
    for all cases listed in Table~\ref{table:BHNS_ID}. The time
    has been shifted by $t_B$, at which time the magnetic field is seeded
    in the NS.
    \label{fig:M0_outside}}
\end{figure}

Consider the binary separation at which the star is tidally
disrupted. It can be estimated by (see Eq.~(17.19) in~\cite{BSBook}) 
\begin{equation}
  R_{\rm tid}\simeq 2.4\,q^{-2/3}\,\mathcal{C}^{-1}\,M_{\rm BH}\,.
\label{eq:r_tidal}
\end{equation}
For a star with compaction $\mathcal{C}=0.145$ and mass ratio $q=3:1$ we find that
disruption  distance is $R_{\rm tid}\simeq 8.0M_{\rm BH}$. On the other hand, we  estimate
the initial position of the ISCO using~Eq.~(2.21) in~\cite{Bardeen72ApJ},
which is strictly correct for a test particle in a Kerr spacetime (see~\cite{TBFS07b} for
a careful analysis). We find that the ISCO  ranges from $R_{\rm ISCO}\sim 7.5M_{\rm BH}$
(for Aliq3sm0.5 case) to $\sim 3.2M_{\rm BH}$ (for Aliq3sp0.75 case). We expect thus heavier
disks in configurations with higher spinning BH.

\begin{figure}
  \centering
  \includegraphics[width=0.49\textwidth]{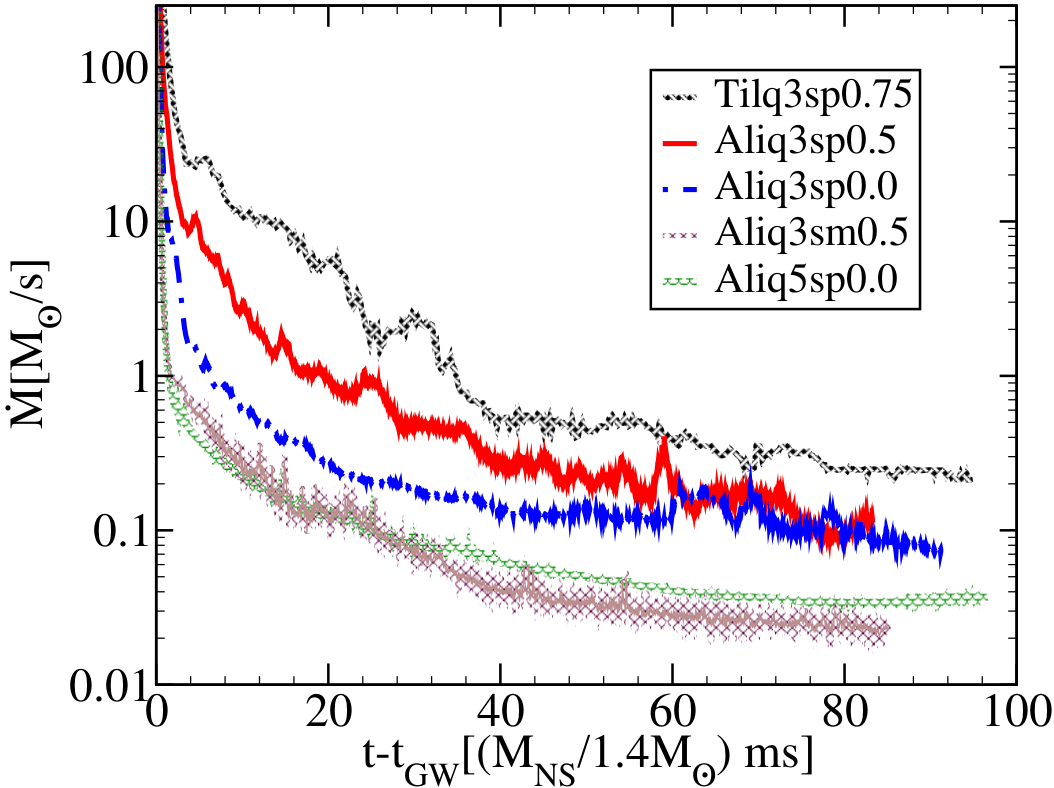}
  \caption{Rest-mass accretion rate for all case listed in Table~\ref{table:BHNS_ID}
    computed via Eq.~(A11) in~\cite{Farris:2009mt}. Time is measured  from the moment
    (retarded time $t-r$) of maximum GW amplitude $t_{\rm GW}$.
    \label{fig:M0_dot}}
\end{figure}

\begin{figure}
  \centering
  \includegraphics[width=0.49\textwidth]{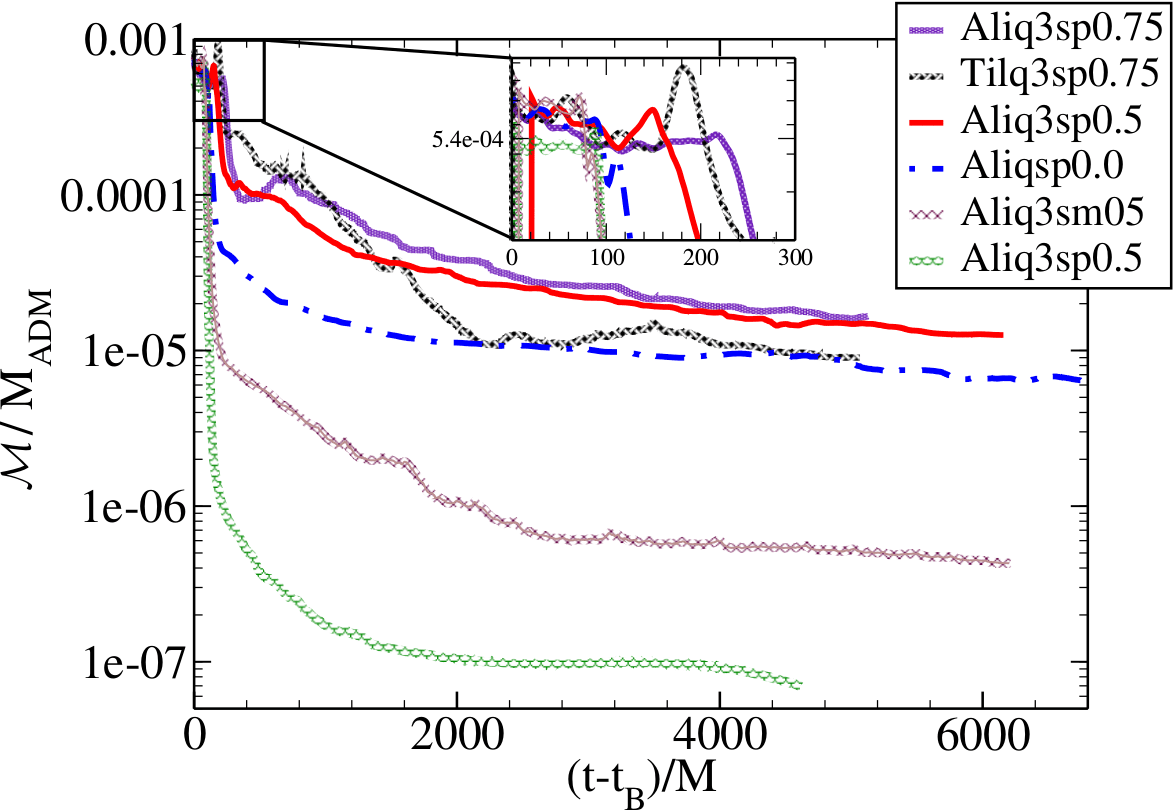}
  \caption{Total magnetic energy $\mathcal{M}$
    outside the BH apparent horizon for all cases listed in Table~\ref{table:BHNS_ID},
    normalized to the ADM mass $M_{\rm ADM}=9.3\times 10^{54}(M_{\rm NS}/1.4M_{\odot})$erg. The
    inset shows that there is no significant enhancement of $\boldmath\mathcal{M}$
    during disruption. The time has been shifted by $t_B$ at which moment the magnetic
    field is seeded in the NS.
    \label{fig:EM_outside}}
\end{figure}

\begin{figure}
  \centering
  \includegraphics[width=0.49\textwidth]{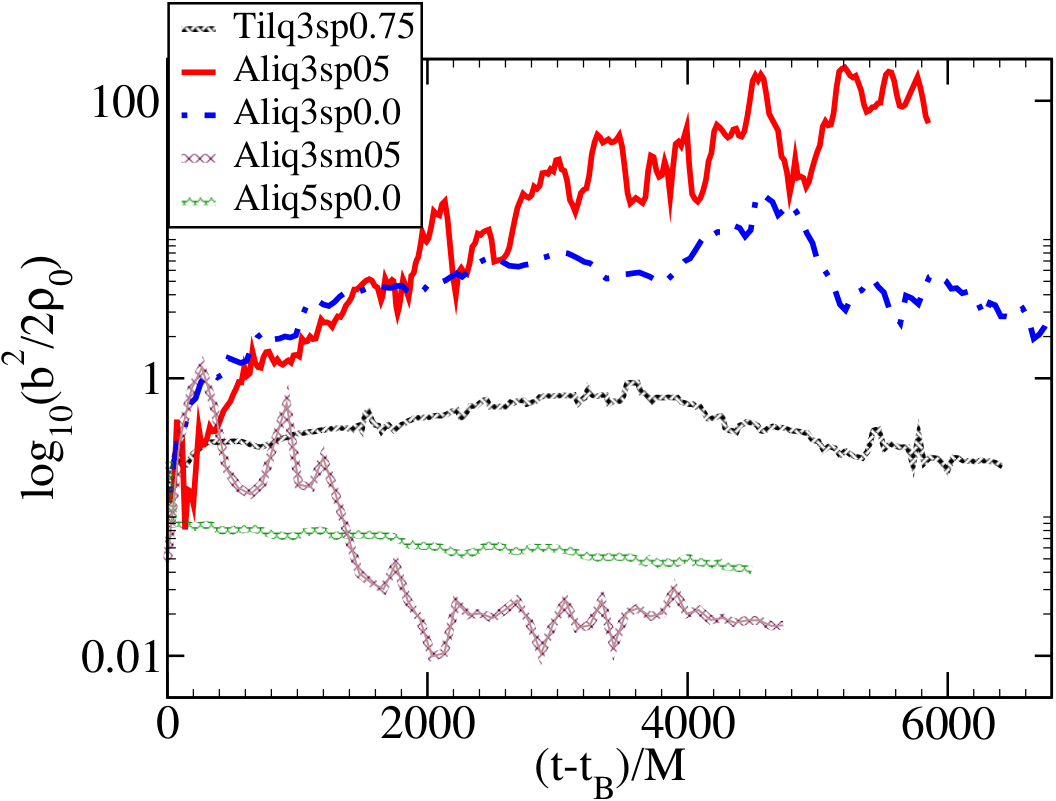}
  \caption{Average value of the force-free parameter $b^2/(2\rho_0)$ vs time (log scale) for
    all cases listed in Table~\ref{table:BHNS_ID}.  The average is computed using grid points
    contained in a cube of edge $2R_{\rm BH}$ above the BH. Here $R_{\rm BH}$
    denotes the radius of the BH apparent horizon.
    \label{fig:b2_2rho0}}
\end{figure}

After $t-t_{B}\sim 40M\sim 1(M_{\rm NS}/1.4M_\odot)$ms following the onset of accretion the bulk
of NS in the case Aliq3sm0.5 is quickly swallowed by the BH companion along with its
frozen-in magnetic field  (see~Fig.~\ref{fig:M0_outside}). Only a tiny fraction of tidally
disrupted debris (less than $1\%$ of the rest-mass of the NS) is left to form the a disk
around a BH remnant with spin~$\tilde{a}\sim 0.3$ (see Table~\ref{table:results_allBHs}).
The rest-mass  accretion  rate computed through Eq.~\ref{eq:Mdot} settles
down to $\dot{M}=8\times 10^{-2}M_\odot/s$ by  $t-t_{\rm GW}\approx 690M\sim 26 (M_{\rm NS}/
1.4M_\odot)\rm ms$ and  then decays slowly (see Fig.~\ref{fig:M0_dot}). Here $t_{\rm GW}$
corresponds to the time (retarded) of  the peak GW amplitude measured at $r_{\rm ext}
\approx 60M\sim 455(M_{\rm NS}/1.4M_\odot)\rm km$. Fig.~\ref{fig:EM_outside} shows the
evolution of the magnetic energy $\mathcal{M}$ outside the BH horizon. During
the first $\sim 40M$ following the onset of the accretion, the magnetic energy plummets
by three orders of magnitude~(see Table~\ref{table:results_allBHs}), as expected. By the
time we terminate the simulation~[$t-t_{\rm GW}\sim 3000M\sim 75(M_{\rm NS}/1.4M_\odot)\rm ms$],
we do not find any evidence of an outflow or tightly wound and globally collimated magnetic
field~(see right top panel in Fig.~\ref{fig:BHNS_case_s0_sm05}), although  we observe that
the field lines just above the BH poles have been partially wound into a helical structure
within~$\sim 2R_{\rm BH}$, due to low density fluid motion. At that time, the rms
value of the magnetic field above the BH pole is only~$\sim 10^{13.3}(1.4M_\odot/M_{\rm NS})\rm G$,
which is expected because only the weakly magnetized external layers of the star survive
the merger and form the disk and the field is not amplified much during the post-merger phase
(see Fig.~\ref{fig:b2_2rho0}). Not surprisingly, {\it a basic ingredient for jet launching
  is a sizable remnant accretion disk.}

%
%

%
%

%
%

%
\begin{figure*}
  \centering
  \includegraphics[width=0.49\textwidth]{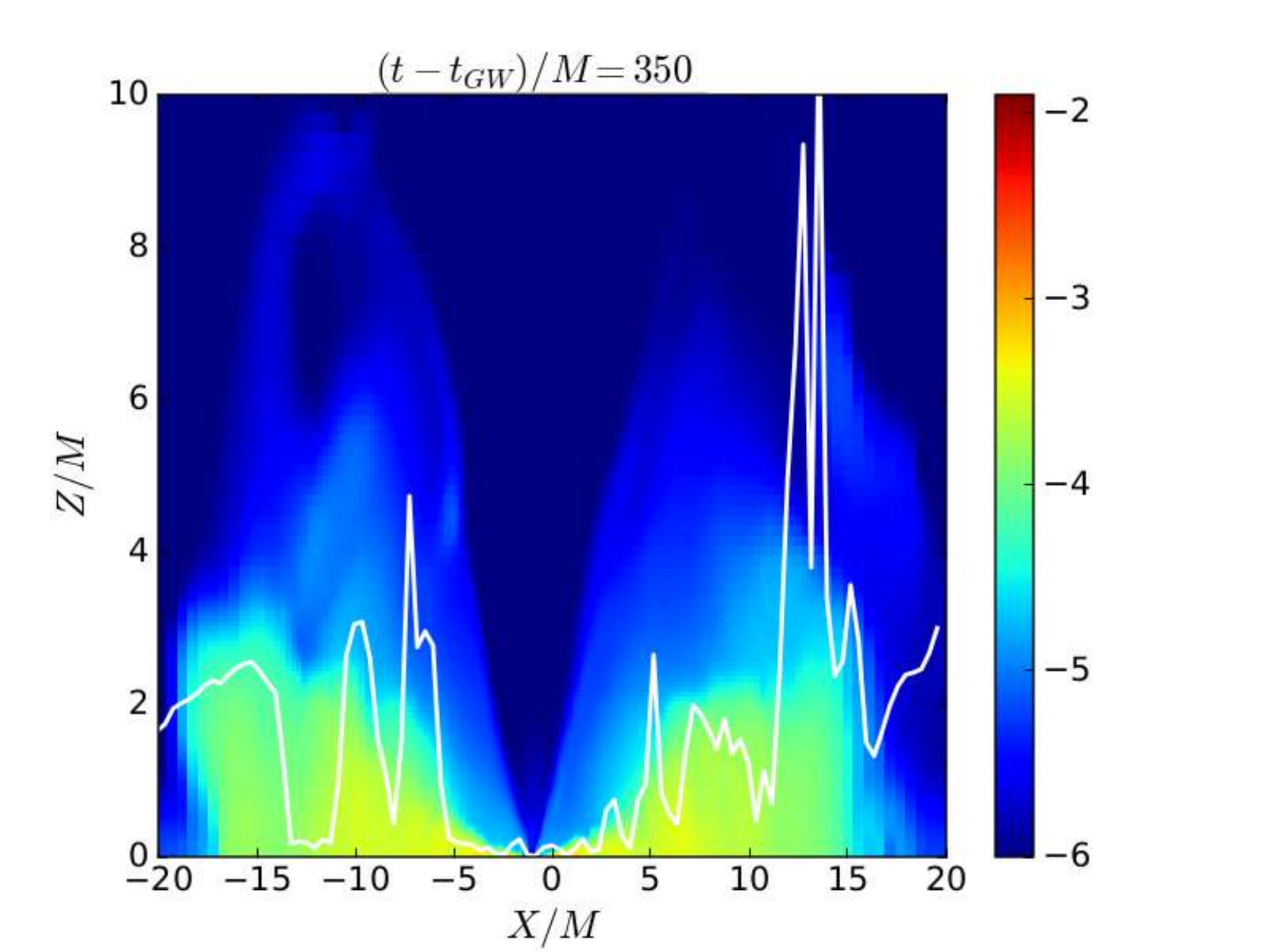}
  \includegraphics[width=0.49\textwidth]{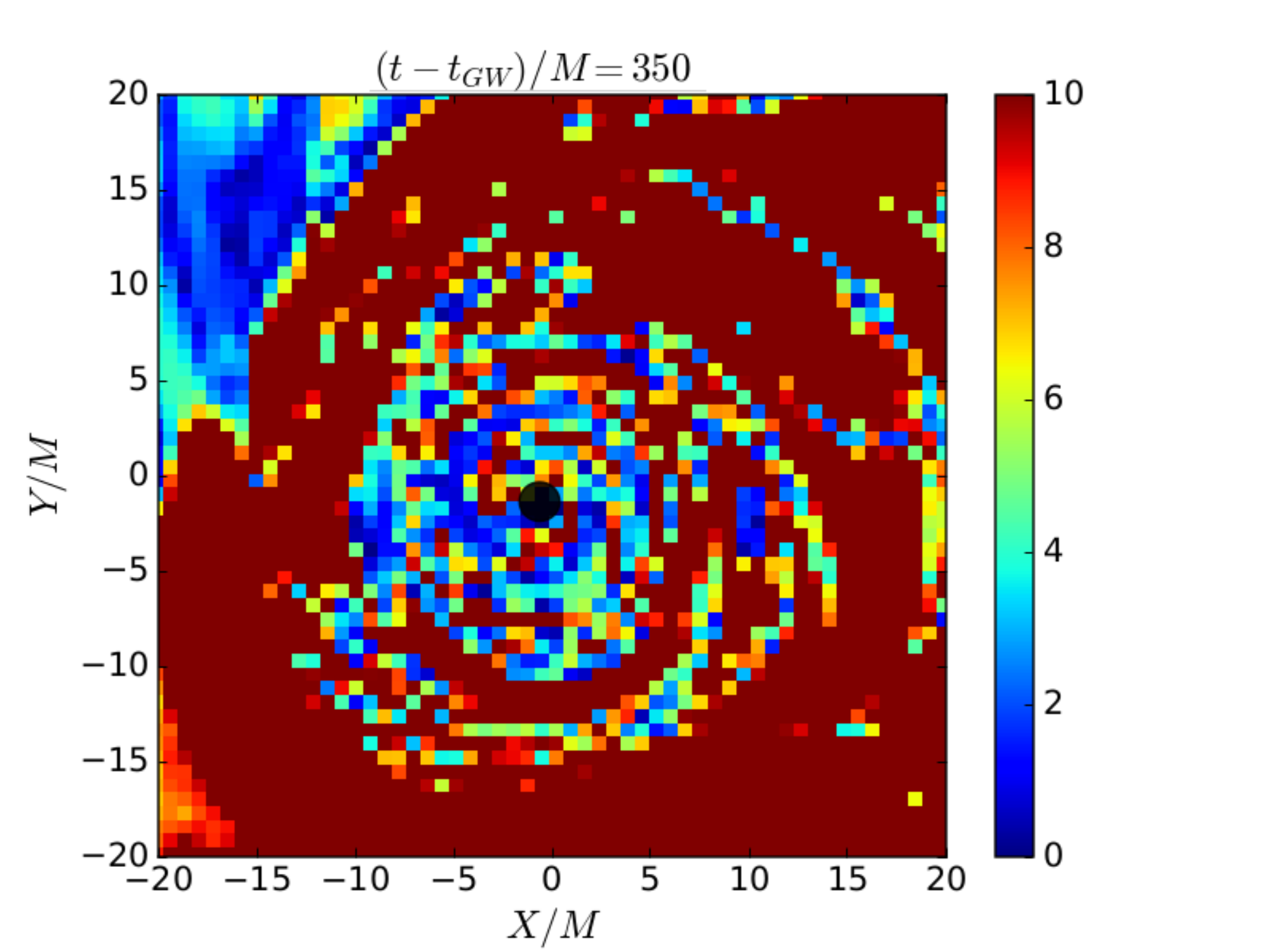}
  \caption{Rest-mass density on the meridional plane along with the $\lambda_{\rm MRI}/2$
    (left panel), and the  quality factor $Q_{\rm MRI}$ on the equatorial plane (right panel)
    at $t-t_{\rm GW}\sim 350M\sim 8.75(M_{\rm NS}/1.4M_\odot)\rm ms$ following the peak GW amplitude
    in case $\tilde{a}=0.5$  (Aliq3sp0.5) but similar behavior among all cases
    with spinning BH $\tilde{a}\geq 0$. The BH apparent horizon is denoted by the black disk. 
    \label{fig:mri}}
\end{figure*}

On the other hand, as the BH spin increases the ISCO shrinks, and therefore the NS 
can be totally disrupted before being swallowed by the BH companion (see right top
and left middle panels in Fig.~\ref{fig:BHNS_case_q_3_1_s05}). The larger the BH spin,
the longer the tidal tails, and thus the heavier the accretion disk (see Table
\ref{table:results_allBHs}). By about  $t\sim 1200M\sim 30(M_{\rm NS}/1.4M_\odot)$ms
following the peak of the accretion ($t-t_B\sim 300M$), the remnant disk settles
with a  mass of $\sim 4.43\%$ of the rest-mass of the NS in case Aliq3sp0.0,~$\sim 10.1\%$
in case Aliq3sp0.5, and $\sim 15.2\%$ in case Aliq3sp0.75 (see Fig.~\ref{fig:M0_outside}),
and then slowly decreases in mass as the accretion proceeds. Similar values were reported
in~\cite{Etienne:2007jg,Etienne:2011ea}, indicating that the seeded magnetic field has a low
impact on the formation of the  disk remnant (see Table~\ref{table:results_allBHs} for
values near the end of the simulations).

By~$t-t_{\rm GW} \approx 1500M\sim 38(M_{\rm NS}/1.4M_\odot)$ms, the rest-mass accretion rate
in the three cases  begins to settle to quasi-equilibrium (see Fig.~\ref{fig:M0_dot}),
and then slowly decays~(see also Table~\ref{table:results_allBHs}). By the time we terminate the
simulations we find that  ${\dot M}\approx(0.09,\,0.12,\, 0.25)M_\odot/s$, for cases $\tilde{a}= 0.0,
\,0.5,\,\rm{and}\,\,0.75$, respectively. The  remnant disk is hence expected to be accreted
in $\Delta t \sim M_{\rm disk}/\dot{M}\sim 0.36(M_{\rm NS}/1.4M_\odot)$s
for Aliq3sp0.0, in $\Delta t \sim 0.75(M_{\rm NS}/1.4M_\odot)$s for Aliq3sp0.5, and
in~$\Delta t \sim 0.5(M_{\rm NS}/1.4M_\odot)$s for Aliq3sp0.75.

During the tidal disruption and  the  early disk + BH phase,  the frozen-in magnetic
field is either stretched and wound into a predominantly toroidal configuration as part of
the tidal tail wraps around the BH forming the accretion disk, or stretched by the low
density material dumped in the atmosphere in the poloidal direction (see right top
and left middle panels in Fig.~\ref{fig:BHNS_case_q_3_1_s05}). However, during those phases we
do not observe a significant enhancement of the total magnetic energy (see Fig.
\ref{fig:EM_outside}) which is expected since initially the magnetic field has an
equipartition--strength [$B_{\rm pole}\simeq 6.7\times 10^{15}(1.4
  M_\odot/M_{\rm NS})$G], i.e. magnetic energy $\approx$ kinetic
energy~\cite{kskstw15}. During $t\sim 40M=1(M_{\rm NS}/1.4M_\odot)\rm ms$  following the
onset of accretion, the bulk of the star, which contains most of the magnetic energy, is
swallowed by the BH (see Fig.~\ref{fig:M0_outside}) leaving only $\sim 6\%$ of the total
initial $\mathcal{M}$ in case Aliq3sp0.0, and $\sim 15\%$ in Aliq3sp0.05 and
Aliq3sp0.75. As the accretion proceeds, the magnetic energy slowly decreases until
quasi--stationary equilibrium is achieved. 

To probe MHD turbulence in the post-merger phase, we compute the effective Shakura--Sunyaev
$\alpha_{\rm SS}$ parameter associated  with viscous dissipation due to magnetic stresses.
In all our cases we find that, between the ISCO and the position of the maximum value of
the rest-mass density, $\alpha_{\rm SS}$ is $\sim 0.01-0.031$ (see Table~\ref{table:results_allBHs}).
Similar values for $\alpha_{\rm SS}$~were found in previous MHD studies of accretion
disks~\cite{Krolik2007,Gold:2013zma}. To check if the MRI is indeed operating in the disk +  BH
remnant, we compute the quality factor $Q_{\rm MRI}$ at $t-t_{\rm GW}\sim 350M\sim 8.75 (M_{\rm NS}/
1.4M_\odot)\rm ms$ following the GW peak amplitude. In the three cases,
we find that in the bulk of the disk  the fastest growing mode of $\lambda_{\rm MRI}$ is resolved by at most
five gridpoints (see Fig.~\ref{fig:mri}), although in some parts it is resolved by more than ten. We
also find that for the most part $\lambda_{\rm MRI}/2$ fits in the disk. As the timescale for MRI is
$\tau_{\rm MRI}\sim \Omega^{-1}\sim 0.1-0.2(M_{\rm NS}/1.4M_\odot)^{1/2}$ms, it is likely that the MRI
is at least partially resolved and operating in the system~\cite{Gold:2013zma}. Here $\Omega$ is the angular
velocity of the disk. The accretion is thus likely driven by MHD turbulence.
%

\begin{figure*}
  \centering
  \includegraphics[width=0.94\textwidth]{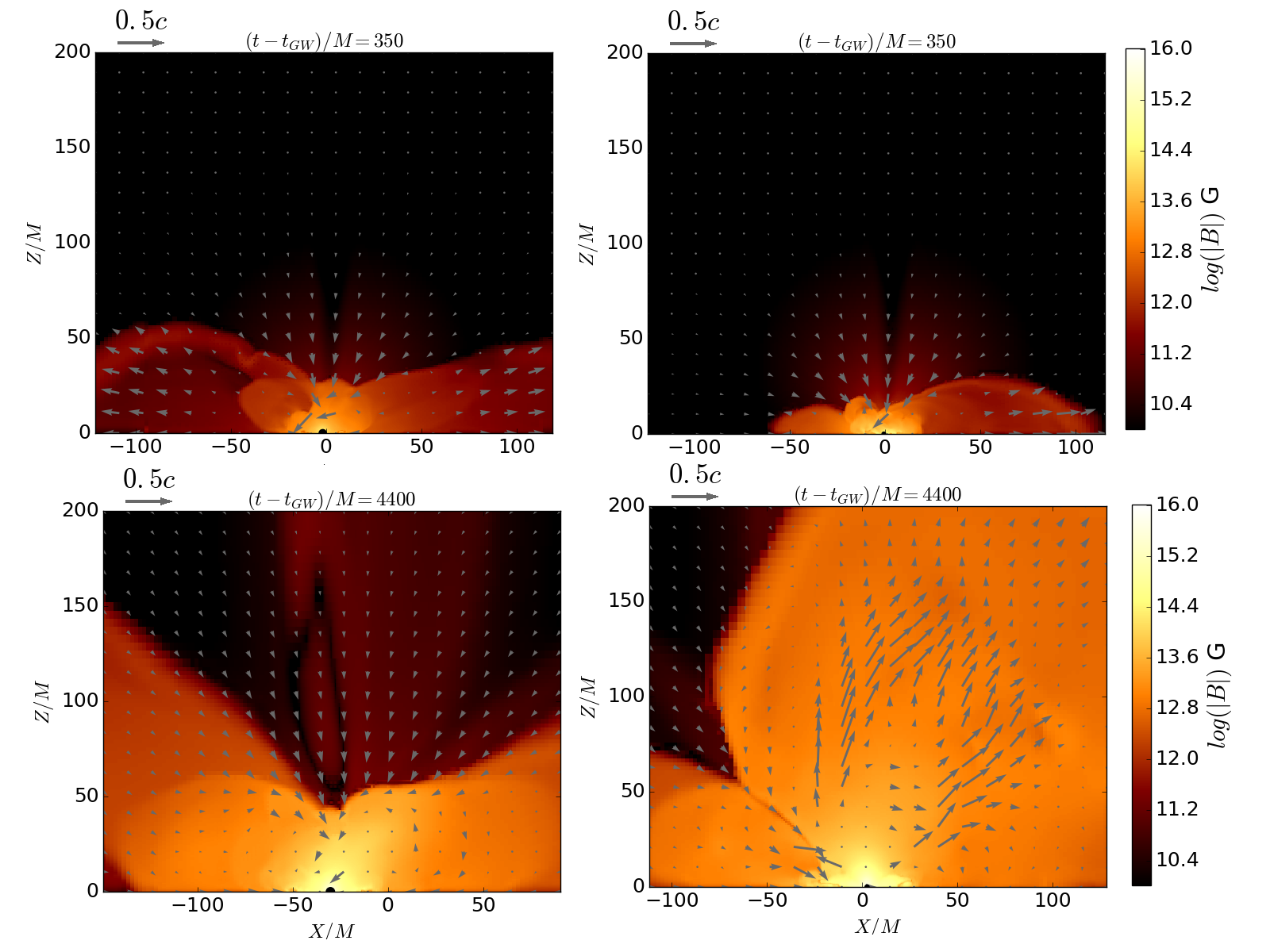}
  \caption{Magnetic field strength on a meridional plane after $t-t_{\rm GW}\sim 350M\sim 8.75(M_{\rm NS}/
    1.4M_\odot)\rm ms$ following
    the maximum GW amplitude, time at which the accretion disk + disk remnant starts to settle, and
    nearly to the end of the evolution for cases Aliq3sp0.0 (left column) and Aliq3sp0.5 (right column).
    Arrows denote the fluid velocity, while the BH apparent horizon is shown as a black disk.
    \label{fig:B_xz_spincase}}
\end{figure*}

Shortly after tidal disruption, the  MRI and magnetic winding  in the disk
convert poloidal to toroidal flux on an Alfv\'en timescale~\cite{Shapiro:2000zh},
$\tau_{\rm A}\sim 1.0 (B/10^{15}\rm G)^{-1}(R_{\rm disk}/50\rm km)
(\rho/10^{14}\rm g/cm^3)^{1/2}\rm ms$, where $R_{\rm disk}$ is the characteristic radius of the
disk (see Eq.~(10.6) in~\cite{shibatabook}), building high magnetic pressure gradients above the
BH and pushing gas outwards above the BH poles
(see top panels in Fig.~\ref{fig:B_xz_spincase}).
As the regions above the BH poles are cleared, the environment  becomes near force-free
($b^2\gg \rho_0$). Depending on the initial spin of the BH companion, we find the following:
%
\begin{figure*}
  \centering
  \includegraphics[width=0.49\textwidth]{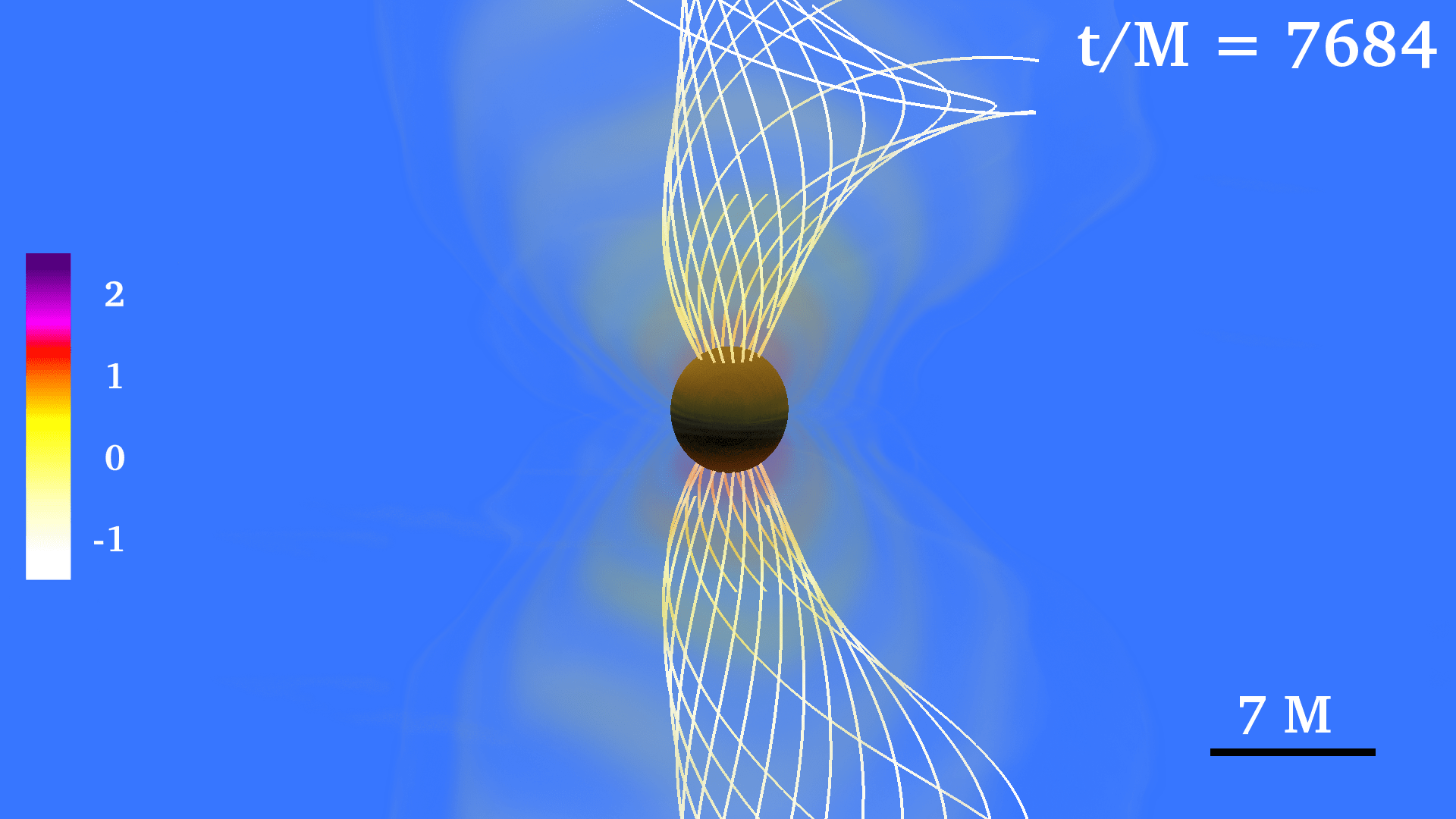}
    \includegraphics[width=0.49\textwidth]{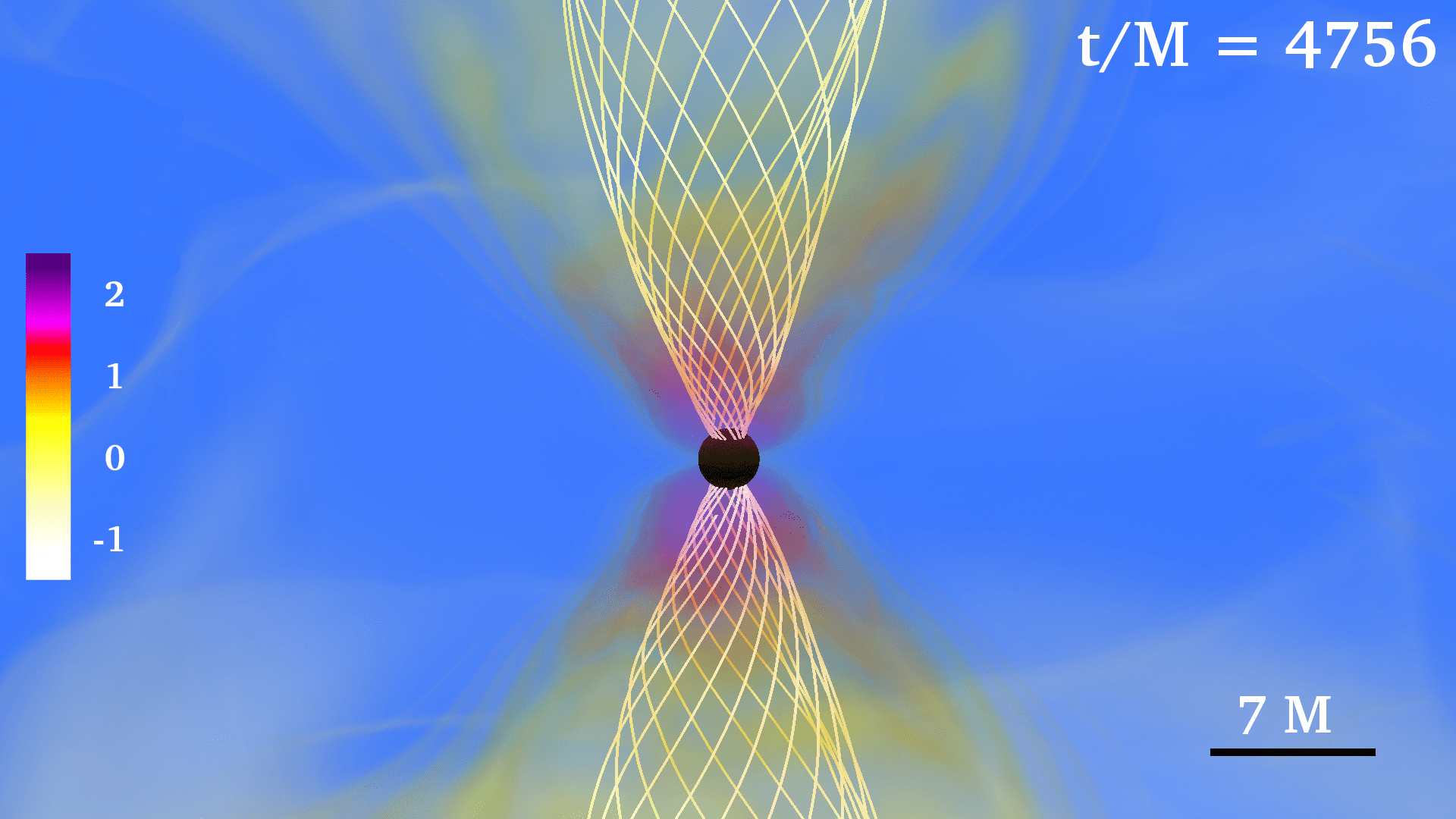}
  \caption{Volume rendering of the ratio $b^2/2\rho_0$ (log scale) near the end
    of the simulation for case Aliq3sp0.0 (left panel) and Aliq3sp0.5 (right panel). The
    magnetic field lines are denoted by white lines plotted in regions where~$b^2/2\rho_0\geq 0$.
    \label{fig:L_angle}}
\end{figure*}
%
\begin{paragraph}{\bf Nonspinning (Aliq3sp0.0) case:}
  By $t-t_{\rm GW}\sim 400M\sim 10(M_{\rm NS}/1.4M_\odot)\rm ms$, we observe that above
  the poles of the BH remnant with spin~$\tilde{a}\sim 0.54$ (see Table~\ref{table:results_allBHs})
  the magnetic field has been wound into a helical funnel  (see middle and right bottom panels in
  Fig.~\ref{fig:BHNS_case_s0_sm05}) but, in contrast with the Aliq3sp0.75 case reported in Paper I,
  there is no evidence of a large-scale sustained outflow. As the magnetic pressure above the BH
  poles increases, magnetically dominated regions ($b^2/2\rho_0^2\gtrsim 1$) expand outwards above
  the BH poles until the magnetic pressure balances the ram pressure produced by fall-back gas at
  a height of~$\sim 15M\sim 115(M_{\rm NS}/1.4M_\odot)\rm km$ (see left bottom in Fig.~\ref{fig:B_xz_spincase}).
  At that height the magnetically dominated regions rise and fall above the BH poles, but
  no longer expand. The left panel in Fig.~\ref{fig:L_angle} shows the magnetically dominated regions
  along with the field lines near the end of the simulation.

  As jet launching via the Blandford--Znajek (BZ) mechanism requires a near force--free  environment
  above the BH poles, we compute the space-averaged  value  of the force-free parameter~$b^2/(2\rho_0)$
  on a cubical region of a
  length  side $2\,R_{\rm BH}$ just above the BH poles during the whole evolution (see~Fig.~\ref{fig:b2_2rho0}).
  We observe that the plasma parameter rapidly grows during the first
  $t-t_{B}\sim 2000M\sim 50(M_{\rm NS}/1.4M_\odot)\rm ms$ following the insertion of the
  magnetic field, and then settles down to~$b^2/2(\rho_{0})_{|ave}~\sim 3$ (see Table
  \ref{table:results_allBHs}).  After about $t-t_B=6000M\sim 150M(M_{\rm NS}/1.4M_\odot)\rm ms$,
  near the end of the simulation, a persistent fall-back flow toward the BH is observed;
  the matter ejected during the disruption has a specific energy $E=-u_0 -1 <0$ (in the asymptotically
  flat region) and eventually rains down with increasing the ram-pressure. However, we also observe
  that magnetic field above the BH poles is amplified from $\sim 10^{13.4}(1.4M_\odot/M_{\rm NS})$G,
  when the disk first settles, to  $\sim 10^{14.6} (1.4M_\odot/M_{\rm NS})$G near the end of the
  simulation (see~Fig.~\ref{fig:B_xz_spincase}).  Hence a longer simulation may  be  needed for a
  magnetically driven outflow to emerge.  However,  if the fall-back debris timescale is longer than
  that of the disk, jet launching may be suppressed. This suggests that {\it there may be a threshold
    value of the initial BH spin below which a sustained outflow is suppressed.}
\end{paragraph}

\begin{paragraph}{\bf Spinning (Aliq3sp0.5 and Aliq3sp0.75) cases:}
  As in the above case, by $t-t_{\rm GW}\sim 400M \sim 10(M_{\rm NS}/1.4M_\odot)\rm ms$ when the remnant
  disk + bh first settles (see bottom panel of Fig.~\ref{fig:L_GW}), the field lines have been wound into
  a helical funnel (see right top  and
  left middle panels in Fig.~\ref{fig:BHNS_case_q_3_1_s05}). However, in contrast to case Aliq3sp0.0,
  as the accretion above the remnant BH poles proceeds, the atmosphere becomes thinner, and the magnetic
  pressure gradients grow. Fig.~\ref{fig:b2_2rho0} shows that following
  the magnetic field insertion, the force-parameter $b^2/2\rho_0$ above the BH poles grows
  from $\sim 10^{-1}$ to $\gtrsim 100$ (see also right panel in Fig.~\ref{fig:L_angle})~near the
  end of the simulation (see Table~\ref{table:results_allBHs}).
  Eventually the magnetic pressure  settles to a value $b^2/2\rho_0\gtrsim 10$
  that allows it to overcome the ram-pressure of the atmosphere. At about  $t-t_{\rm GW}\sim 400M\sim
  10(M_{\rm NS}/1.4M_\odot)$ms, the inflow is halted, and a magnetically sustained outflow 
  emerges~(see bottom panels in Fig.~\ref{fig:BHNS_case_q_3_1_s05}). The unbound outflow ($E=-u_0-1>0$)
  extends to heights greater than $100 M\sim 760(M_{\rm NS}/1.4 M_\odot)$km in Aliq3sp0.5
  ($\tilde{a}=0.5$) at  $t-t_{\rm GW}\sim 3500{\rm M}\sim 88(M_{\rm NS}/1.4M_\odot)$ms, and at $t-t_{\rm GW}\sim
  4000{\rm M}\sim 100(M_{\rm NS}/1.4M_\odot)$ms in~Aliq3sp0.75 ($\tilde{a}=0.75$).
   The characteristic maximum value of the Lorentz factor in the funnel is $\Gamma_L\sim 1.2-1.3$.
  So, we conclude that by $\gtrsim 88(M_{\rm NS}/1.4 M_\odot)$ms these two cases launch an incipient
  jet --an unbound and mildly relativistic outflow within a tightly wound, collimated, helical magnetic
  funnel above the BH poles. The delay of the jet launching  in Aliq3sp0.75 with respect to that
  in  Aliq3sp0.5  is likely due to a heavier atmosphere; a larger ejection of the matter  outside the
  ISCO occurs for higher spins.   Although the jet is only mildly relativistic, it is expected that
  the jet will be accelerated to $\Gamma_L\gtrsim 100$ as required by sGRB models. As it was pointed out
  in~Paper I, the maximum attainable Lorentz factor of a magnetically--powered, axisymmetric jet is
  $\Gamma_L^{\rm max}\sim b^2/2\rho_0$~\cite{Vlahakis2003}. The lifetime of the engine fuel (lifetime
  of the disk) is $\Delta t\sim 0.5-0.75(M_{\rm NS}/1.4M_\odot)s$ and thus consistent with
  sGRBs~\cite{Bhat:2016odd}.  We  also observe a magnetic field amplification  above the BH poles
  from $\sim 10^{13.4}(1.4M_\odot/M_{\rm NS})$G, when the disk first settles, to  $\gtrsim 10^{15}
  (1.4M_\odot/M_{\rm NS})$G near the end of the simulation
  (see right bottom panel in Fig.~\ref{fig:B_xz_spincase}).

  The level of collimation of the jet is measured by the  funnel  opening  angle $\theta_{\rm jet}$, which
  is defined as polar angle at which  the Poynting flux drops to $50\%$ of its maximum. Based on the
  angle distribution of the outgoing flux on the surface of a sphere with coordinate radius $60M\sim
  460(M_{\rm NS}/1.4M_\odot)\rm km$  (see Fig.~\ref{fig:L_angleII}), we estimate that the opening angle
  of  the jet is $\sim 25^\circ-30^\circ$. 

  We compute the ejecta via $M_{\rm esc} =\int_{|u_t|>1} \rho_0 d^3x$
  at different radii between $30M\sim 230(M_{\rm NS}/1.4M_\odot)\rm km$ and $100M\sim 760(M_{\rm NS}/
  1.4M_\odot)\rm km$. We find that in these cases the rest-mass fraction $M_{\rm esc}/M_{NS}$ of the escaping mass
  is $\sim 10^{-2}$, and thus in principle could be detected with the Large Synoptic Survey Telescope 
  \cite{Metzger:2011bv} and give rise to Kilonovae phenomena~\cite{Metzger:2016pju}.

%
%
\begin{table*}[th]
  \caption{Comparison of simulation results with the unified model presented in~\cite{Shapiro:2017cny}.
    \label{table:Tmodel}}
  \begin{center}
	\begin{tabular}{ccccccccc} 
	  \hline         
	  \hline
        \multirow{2}{*}{Case}  &
      \multicolumn{2}{c}{\underline{\hspace{0.25cm} $L_{\rm jet}\,\rm (erg/s)$\hspace{0.25cm}} } &
      \multicolumn{2}{c}{\underline{\hspace{0.25cm} $\dot{M}_{\rm BH}\,(M_\odot/s)$\hspace{0.25cm}} } &
      \multicolumn{2}{c}{\underline{\hspace{0.25cm} $\rho\,[(1.4M_\odot/M_{\rm NS})^2]\,(\rm g/cm^3)$\hspace{0.25cm}}} &
      \multicolumn{2}{c}{\underline{\hspace{0.25cm} $B_p\,[(1.4M_\odot/M_{\rm NS})]\,(\rm G)$ \hspace{0.25cm}}}             \\
	 & Model    &  Simulations &Model & Simulations & Model & Simulations & Model  & Simulations \\
	\hline
	  Aliq3sp0.5  & $10^{52}$   &$10^{52}$   & $10^0$  & $10^{-1}$ & $10^{10}$  & $10^{9}$  & $10^{16}$ & $ 10^{15}$ \\
	  Aliq3sp0.75 & $10^{52}$   &$10^{51}$   & $10^0$  & $10^{-1}$ & $10^{10}$  & $10^{10}$ & $10^{16}$ & $ 10^{15}$ \\
	\hline
	\end{tabular}
  \end{center}
\end{table*}

To further assess if the BZ mechanism~\citep{BZeffect} is operating
in our BHNS remnants, we compute the ratio of the angular velocity of the magnetic field 
$\Omega_F\equiv F_{t\theta}/F_{\theta\phi}$  to the angular velocity of the BH defined as
  \begin{equation}
    \Omega_H=\frac{\tilde a}{2M_{\rm BH}}\, \left(1+\sqrt{1-\tilde a^2}\right)\,,
  \end{equation}
on a meridional plane passing through the BH centroid and along a coordinate semicircle of
radius $R_{\rm BH}<R_{\rm ext}<2\,R_{\rm BH}$ as in~Paper~I. Here  $F_{\mu\nu}$ is the Faraday
tensor. Notice that the definition  of $\Omega_F$ is strictly valid for stationary and axisymmetric
spacetimes in Killing coordinates~\cite{Blandford1977}. In both cases we find that the ratio
$\Omega_F/\Omega_H$ ranges from $\sim 0.4-0.45$ at the BH pole to $\sim 0.1$ near the equator.
The deviation from the
  expected split-monopole value $\Omega_F/\Omega_H\sim 0.5$ (see~\cite{Komissarov2001})
  can be attributed to the deviations from  a split-monopole magnetic field, the gauge
  in which $\Omega_F$ is computed, and/or inadequate resolution. On the other hand, the
  outgoing Poynting luminosity is $L_{\rm jet}\sim 10^{51.2}-10^{51.6}$ (see top panel
  of Fig.~\ref{fig:L_GW}), which is consistent with  that generated  by  the
  BZ mechanism~\cite{Thorne86}
 \begin{equation}
   L_{\rm BZ}\sim10^{51}\,\tilde{a}^2\,\left(\frac{M_{\rm BH}}{5.6M_\odot}\right)^2\,
   \left(\frac{B}{10^{15}\rm G}\right)^2 \rm erg\,/ s\,.
 \end{equation}
 It is therefore likely that the BZ  mechanism is operating in our systems.
 Note that we normalized the mass of the BH to $5.6M_\odot$ because 
   $\gtrsim 90\%$ of the rest-mass of the NS is swallowed by the BH during merger
   (see Table~\ref{table:results_allBHs}).
 
 In contrast to cases Aliq3sm0.5 and Aliq3sp0.0, the  BHNS configurations Aliq3sp0.5 and Aliq3sp0.75
 launch a mildly relativistic outflow sustainable by a helical magnetic field. These results suggest that
 {\it the ingredients for jet launching from the remnant of BHNS mergers are: (1) a binary
   companion that contains a spinning BH (for sizable disks)}, and {\it (2) a strong NS poloidal exterior
   magnetic field component that ties fluid elements in the disk to
   low density debris above the BH poles.}
 
\subsection{Effect of varying the mass ratio (case Aliq5sp0.0)}
\label{subsec:massratio}
As it can be seen from Eq.~\ref{eq:r_tidal}, the tidal disruption  distance decreases as the
mass ratio of the binary increases. The closer the tidal distance to the ISCO, the smaller the
tidal effect and hence  the smaller the mass of the remnant disk and, consequently, the less
magnetic energy left to launch a jet. The tidal separation for a BHNS configuration with
mass ratio $q=5:1$,  a star compaction $\mathcal{C}=0.145$, and  a  nonspining BH companion
is $R_{\rm tid}\sim 6\,M_{\rm BH}$, which ``coincides'' with the ISCO.

Fig~\ref{fig:BHNS_case_q_5_1_s0} summarizes the evolution  of this case starting from
the insertion of the magnetic field~(left panel), through the tidal disruption and merger
(middle panel), and finally showing the outcome  once the disk + BH remnant relaxes to a
quasi--steady state (right panel). As  expected, the star is somewhat disrupted before it plunges
into the BH. Fig.~\ref{fig:M0_outside} shows that during the first $t-t_{B}\sim 40M\sim 1.5(M_{\rm NS}
/1.4M_\odot)\rm ms$  following the onset of accretion the bulk of NS is quickly swallowed leaving an
``orphan'' BH remnant surrounded by a small, weakly magnetized cloud (less than $1\%$ of the
rest-masss of the star) to form the accretion disk. By  $t-t_{\rm GW}
\approx 680M\sim 26(M_{\rm NS}/1.4M_\odot)\rm ms$ the rest-mass accretion  rate settles
down to~$\dot{M}=1.4\times 10^{-2}M_\odot/s$  and  then decays slowly (see Fig.~\ref{fig:M0_dot}).
Fig.~\ref{fig:EM_outside} clearly shows that during that period there is basically no
magnetic energy left (see Table~\ref{table:results_allBHs}) as the frozen-in magnetic field has
been dragged into the BH during the plunge phase.  We do not find evidence of magnetic
field collimation or an outflow. Near to the end of the simulation the magnetic field strength
above the BH poles is $\lesssim 10^{12.3}(1.4M_\odot/M_{\rm NS})$G. 
%
\begin{figure}
  \centering
  \includegraphics[width=0.48\textwidth]{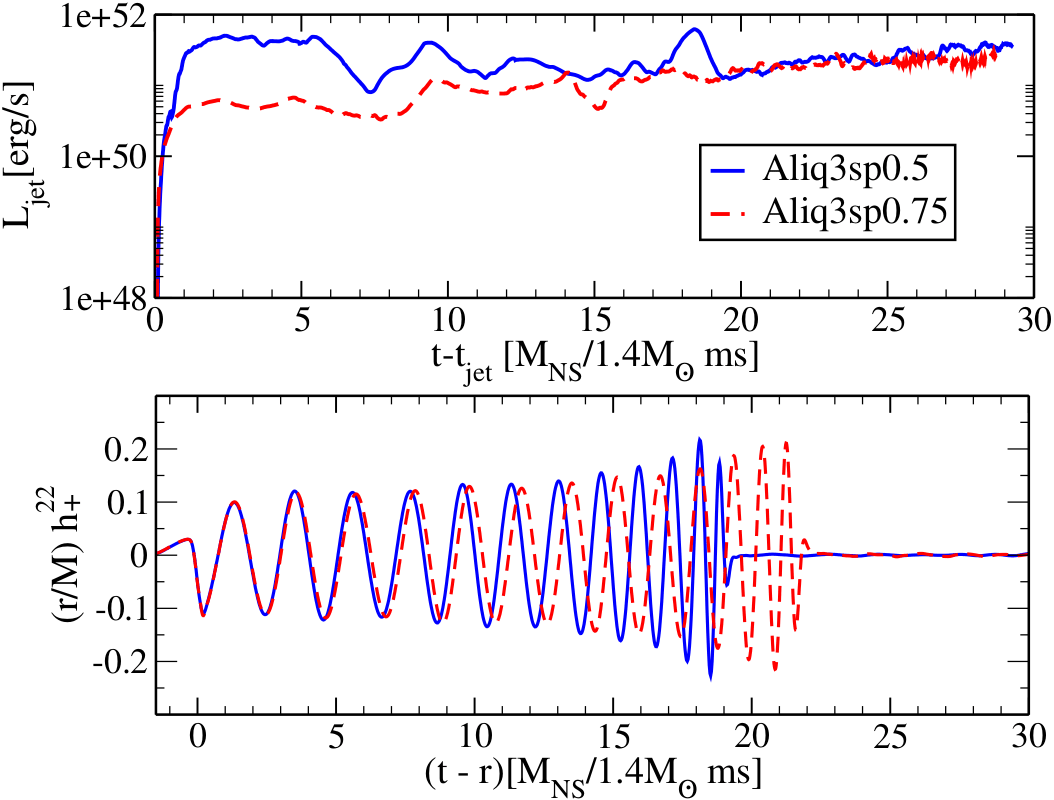}
  \caption{Outgoing EM (Poynting) luminosity for $t\geq t_{\rm jet}$ computed at
    a  coordinate  sphere  of  radius $r=100M\sim 760(M_{\rm NS}/1.4M_\odot)\rm km$
    (top panel), and (2,2) mode of the  gravitational wave strain $h_+$  as functions of
    retarded time extracted at $r_{\rm ex}=80M\sim 606(M_{\rm NS}/1.4M_\odot)\rm km$
    (bottom panel) for case Aliq3sp0.5 (continuous line) and case Aliq3sp0.75 (dashed line).    
    \label{fig:L_GW}}
\end{figure}
\end{paragraph}
%
\begin{figure}
  \centering
  \includegraphics[width=0.48\textwidth]{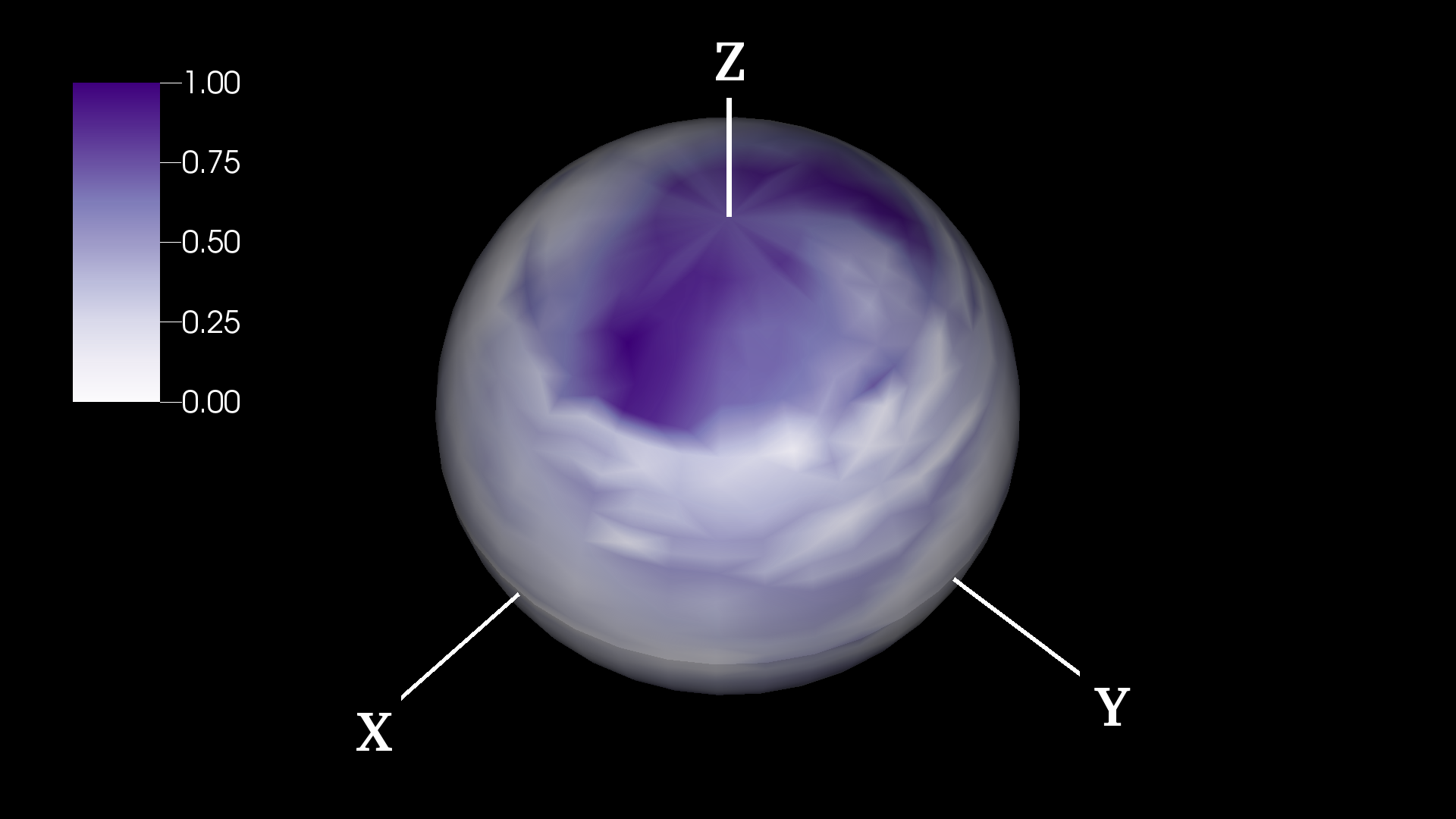}
  \caption{Angular distribution of Poynting flux for case $\tilde{a}=0.5$, normalized by
    its peak value on a sphere of radius  $60M= 4600(M_{NS}/1.4M_\odot)$km.
    Angles are defined with respect to a spherical coordinate system
        centered on the BH center, with the spin axis along the $z$ direction.
    \label{fig:L_angleII}}
\end{figure}

Notice that population synthesis studies have suggested that the most likely BHNS mass ratio may
be $q=7:1$~\cite{Belczynski:2007xg,bdbofh10}, although recently it has been suggested  how
low-mass BH formation channels may arise in BHNS~\cite{Yang:2017gfb}. For this high mass ratio
configuration with a typical NS of compaction $\mathcal{C}=0.145$, the binary
tidal separation is $R_{\rm tid}\sim 0.45M_{\rm BH}$. So, the critical spin at which tidal
disruption occurs at the ISCO is $\tilde{a}=0.375$. As the basic ingredient for jet launching
is a sizable magnetized disk, the above estimation  suggests that {\it high mass ratio BHNS
configurations may be the progenitors of central engines that power sGRBs only if the spin of the
BH companion is  $\tilde{a}> 0.4$}~(see~also~\cite{Kyutoku:2011vz,Foucart:2018rjc,Foucart:2012nc}.

\subsection{Effect of magnetic field orientation (case Tilq3sp0.75)}
In the above section, we described the effects of the BH spin and mass ratio on the
emergence of an incipient jet when the pulsar-like magnetic field seeded in the NS is aligned
with the total orbital angular momentum of the  system. In the following, we consider a BHNS
configuration in which the BH companion has a spin of $\tilde{a}=0.75$, and the star is seeded
with a pulsar-like magnetic field whose dipole magnetic moment is now  tilted $90^\circ$ with
respect to the orbital angular momentum (see left panel in Fig.~\ref{fig:BHNS_case_q_3_1_s075_tilted}).

The dynamics of the gas during tidal disruption, merger and early disk + BH phases are similar
to those reported in Paper I, and summarized in~Sec.~\ref{subsec:Ef_spin}. This is not unexpected
since the strength of the dynamical unimportant magnetic field in both cases is the same. However,
by around $t-t_{\rm GW}=1600M\sim 40(M_{\rm NS}/1.4M_\odot)$ms, by which time the accretion rate $\dot{M}$
settles down~(see Fig.~\ref{fig:M0_dot}),  the frozen-in magnetic field has been driven into a
predominantly toroidal configuration in the disk, while in the atmosphere, in contrast to the spinning
cases reported in Sec~\ref{subsec:Ef_spin} (see also Paper I), there is no a coherent poloidal magnetic
field configuration~(see right panel in~Fig.~\ref{fig:BHNS_case_q_3_1_s075_tilted}). After evolving
the remnant disk +  BH for $t-t_{\rm GW}\gtrsim 4000M\sim 100(M_{\rm NS}/1.4M_\odot)$ms, we do not
find any evidence of magnetic field collimation or an outflow above the BH poles. As before, we
compute the space-averaged  value  of the force-free parameter~$b^2/(2\rho_0)$ on a cubical region
of a length  side $2\,R_{\rm BH}$ just above the BH poles along the whole evolution (see~Fig.
\ref{fig:b2_2rho0}). Following disruption, we observe that the plasma parameter peaks at two times
its initial value and then slowly decreases until it falls to a value of~$b^2/(2\,\rho_{0})_{|ave}~\sim 0.26$
(see Table~\ref{table:results_allBHs}). After about $t-t_{\rm GW}= 5000M\sim 125M(M_{\rm NS}/
1.4M_\odot)\rm ms$ a persistent fall-back material toward the BH is observed.

When the magnetic field is aligned with the total angular momentum of the system,
vertical field lines thread the BH prior to tidal disruption (see left top panel in Fig.~1 in Paper I).
After disruption, these lines connect the polar regions of the BH to low-density debris in the atmosphere.
Similarly,  fluid  elements in the disk are linked to other fluid elements in the disk, and to those
ejected during the disruption, through external vertical magnetic lines~(see right top panel in Fig.~1 in
Paper I). These two effects induce a strong poloidal magnetic field in the BHNS remnant. By contrast,
in the tilted case Tilq3sp0.75,  horizontal field lines mainly thread the BH prior to tidal disruption
(see left panel~in Fig.~\ref{fig:BHNS_case_q_3_1_s075_tilted}). After disruption, these lines can only
connect the BH poles to the inner part of the new--born disk, and they are rapidly  wound to a predominantly
toroidal configuration. Also, fluid  elements in the disk are linked to other fluid elements in the
disk, and to the low-density debris in the atmosphere, through external predominantly horizontal field
lines. The BHNS remnant hence lacks a coherent poloidal magnetic field component (see right panel in Fig.
\ref{fig:BHNS_case_q_3_1_s075_tilted}).

While the properties of the disk + bh remnant, such as BH spin, mass, and accretion rate, are approximately
independent of the magnetic field topology (see~Table~\ref{table:results_allBHs}),  the emergence of the
jet seems to be very sensitive to it. As it was pointed out in~\cite{GRMHD_Jets_Req_Strong_Pol_fields},
a poloidal magnetic field component with a consistent sign in the vertical direction is required to launch
and support a jet.

The above  results indicate that {\it there is a threshold value of the tilt angle of the dipole magnetic
  moment with respect to the orbital angular momentum below which the poloidal dipole magnetic field
component is suppressed, and with it the emergence of a jet}.

\subsection{Universal model}
\label{subsec:Model}
Recently we proposed a ``universal'' analytic model in~\cite{Shapiro:2017cny} that estimates a number of
global parameters that characterize disk +  BH  remnants that launch jets following BHNS mergers,
BHBH mergers immersed in magnetized disks, and the collapse of massive stars. The jets
are powered by the BZ mechanism and the parameters are determined by only a couple of nondimensional
ratios characterizing the remnant system. This model predicts  the characteristic density in
the accretion disk, the strength of the magnetic field above the BH poles, the  rest-mass accretion
rate after the system has reached a quasi-stationary  state, and most significatively the EM 
(Poynting) luminosity as follows (see~Eqs. 11-13~in~\cite{Shapiro:2017cny}):
\begin{equation}
  \rho\,M_{\rm BH}^2\sim \frac{1}{\pi}\,\left(\frac{M_{disk}}{M_{\rm BH}}\right)\,
  \left(\frac{M}{R_{disk}}\right)^3\,,
\end{equation}
\begin{equation}
  B_p^2\,M_{\rm BH}^2\sim 8\,\left(\frac{M_{disk}}{M_{\rm BH}}\right)\,
  \left(\frac{M}{R_{disk}}\right)^3\,,
\end{equation}
\begin{equation}
  \dot{M}_{\text{eq}}\sim 4\,\left(\frac{M_{disk}}{M_{\rm BH}}\right)\,
  \left(\frac{M}{R_{disk}}\right)^3\,[\dot{\mathcal{M}}_0]\,,
\end{equation}
\begin{equation}
  L_{BZ}\sim \frac{1}{10}\,\left(\frac{M_{disk}}{M_{\rm BH}}\right)\,
  \left(\frac{M_{\rm BH}}{R_{disk}}\right)^3\,\left(\frac{a}{M_{\rm BH}}\right)^{2}\,
       [\mathcal{L}_0]\,,
\end{equation}
where $\mathcal{L}_0\equiv c^5/G = 3.6 \times 10^{59} \rm erg/ s^{}$ and $\dot{\mathcal{M}}_0
\equiv c^3/G = 2.0 \times 10^5 M_{\odot}/\rm s$.  Table~\ref{table:Tmodel} shows a
comparison of our simulations results with the model predictions, i.e. using as input the data in
Table~\ref{table:results_allBHs} to calculate the nondimensional ratios. We find that within an
order of magnitude, the results are consistent. As was pointed out in~\cite{Shapiro:2017cny},
while there exist  different formation scenarios for forming disk + BH systems, and their
disk masses, densities and magnetic field strength vary by orders of magnitude, these features
conspire to generate jet Poynting luminosities that all lie in the narrow range of $10^{52\pm 1}\rm
erg/s$. Interestingly, these luminosity distributions mainly reside in the same narrow  range
characterizing the observed luminosity distributions of over 400 short and long
GRBs~\cite{Li:2016pes}.
%
\begin{figure*}
  \centering
  \includegraphics[width=0.33\textwidth]{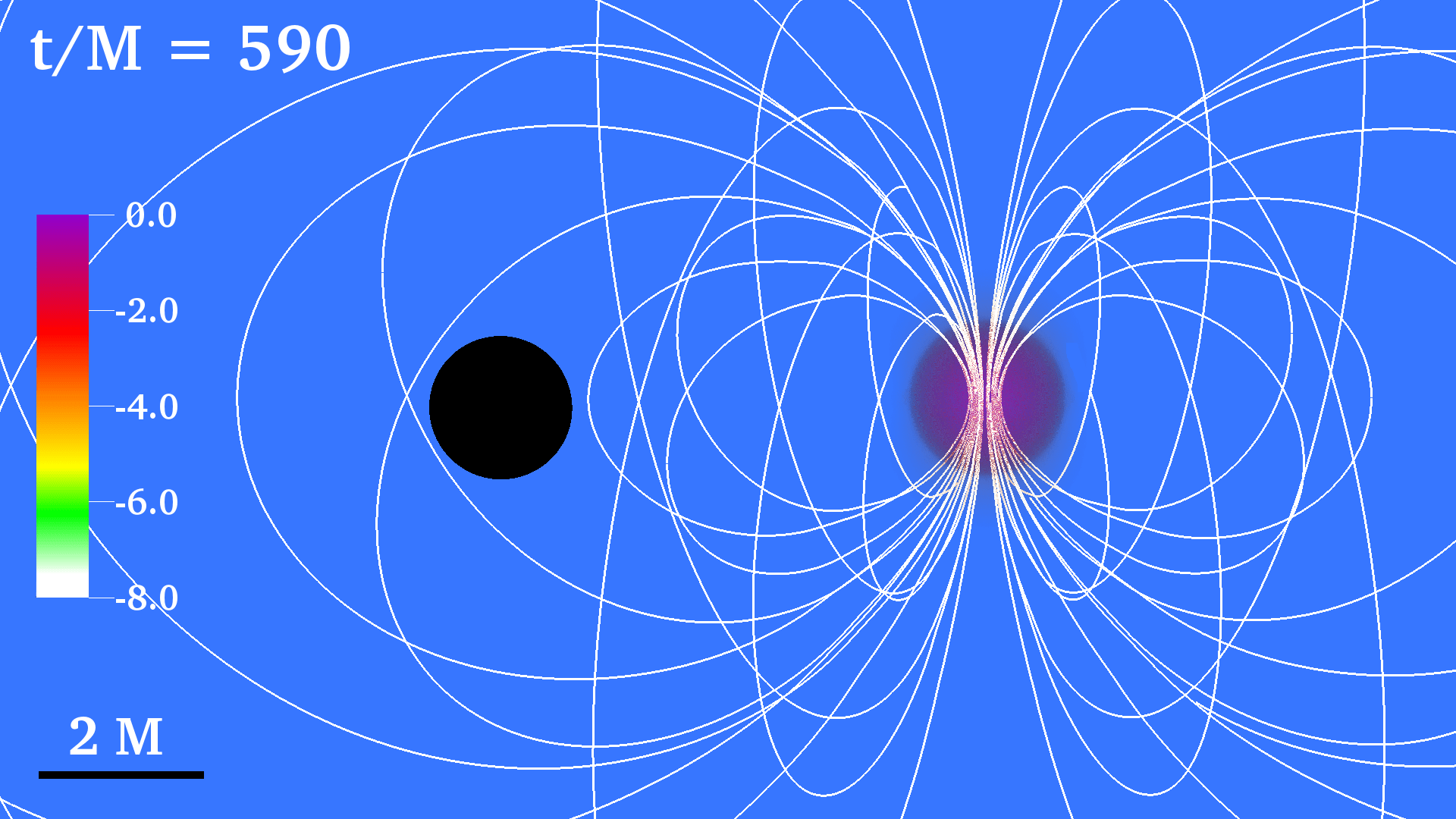}
  \includegraphics[width=0.33\textwidth]{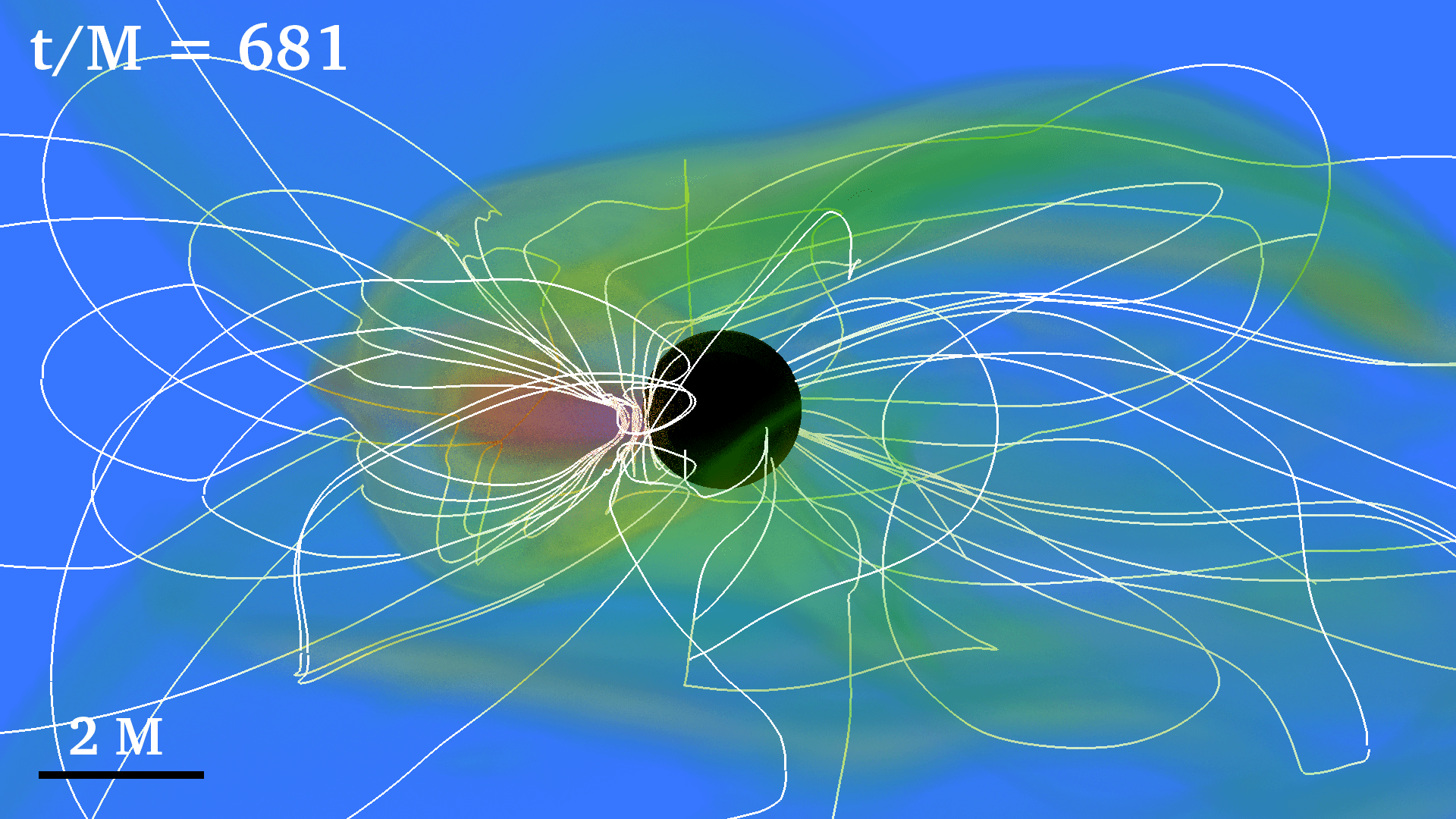}
  \includegraphics[width=0.33\textwidth]{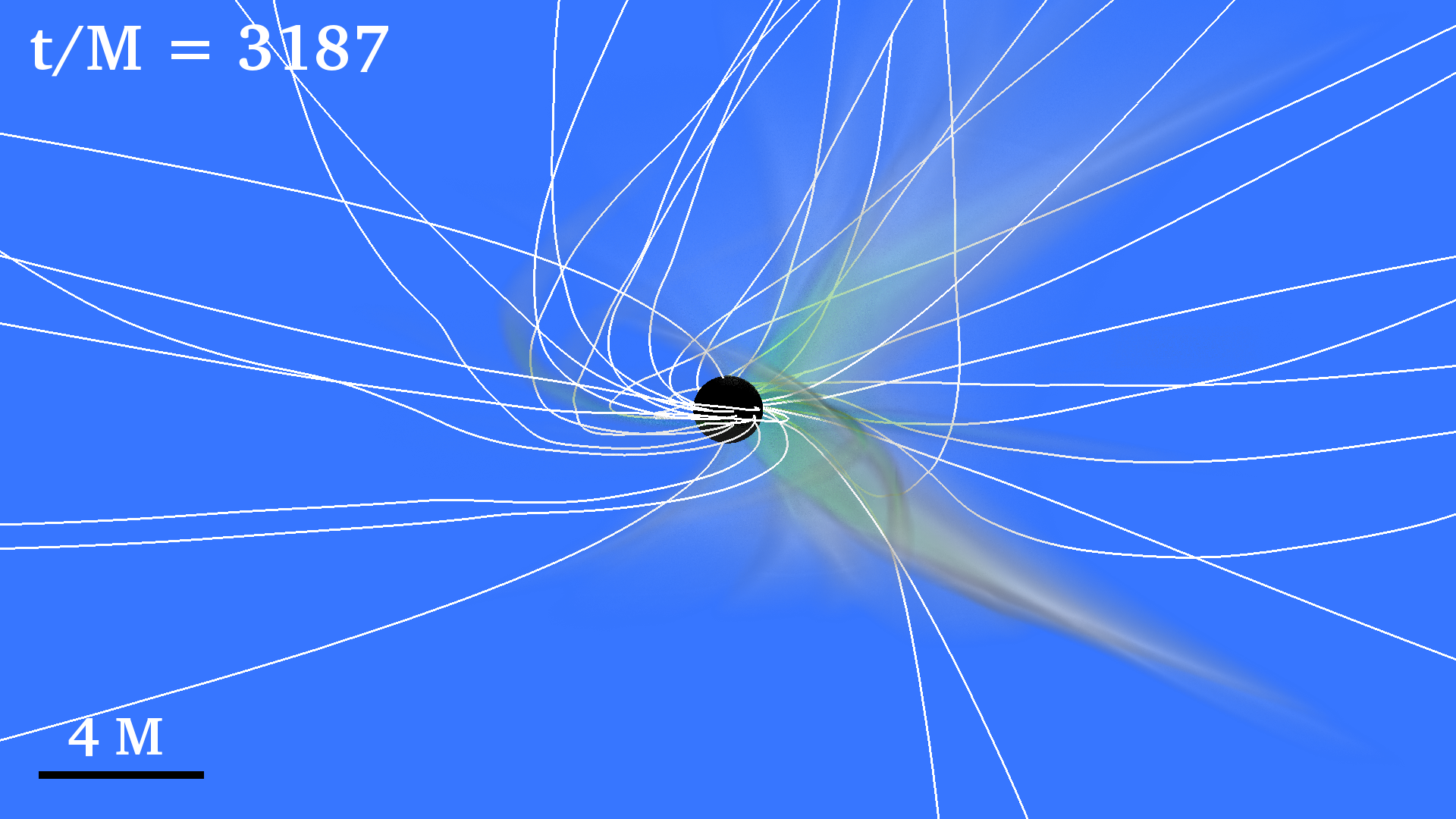}
  \caption{Volume rendering of rest-mass density $\rho_0$ normalized to its initial
    NS maximum value  $\rho_0=8.92\times 10^{14}\,(1.4M_\odot/M_{\rm NS})^2\rm{g\,/cm}^{3}$
    (log scale) at selected times for case  Aliq5sp0.0 (see Table~\ref{table:BHNS_ID}).
    White lines denote the magnetic field while the arrows denote the fluid velocity.
    The BH apparent horizon is denoted as a black sphere.
    Here $M=11.44(M_{\rm NS}/1.4M_\odot)\rm km=3.81\times
    10^{-2}(M_{\rm NS}/1.4M_\odot)\rm ms$.
    \label{fig:BHNS_case_q_5_1_s0}} 
\end{figure*}
%
\section{Conclusions}
\label{sec:conclusion}
The coincident detection of gravitational radiation (event GW170817) with short gamma
ray bursts (GRB 170817A),  detected~$\sim 1.7\rm s$ after the inferred binary merger time
\cite{GBM:2017lvd}, confirm that merging compact binaries, containing at least one neutron
star, can be the progenitors of the engine that powers sGRBs as proposed
by~\cite{Pac86ApJ,EiLiPiSc,NaPaPi}. 
This single multimessenger detection has been already
used to impose some constraints on the maximum mass of a spherical neutron star~\cite{Margalit:2017dij,
  Shibata:2017xdx,Ruiz:2017due,Rezzolla:2017aly}, on the tidal deformability, on the radius of
the star~\cite{Most:2018hfd,TheLIGOScientific:2017qsa, Abbott:2018exr,Radice:2017lry,
  Bauswein:2017vtn}, and other properties of the progenitor stars.

We recently reported the first self-consistent numerical calculations in full GR that demonstrate
that the remnant of magnetized BHNS mergers can launch an incipient jet
if the star  is initially seeded with a dipole magnetic field that extends from the NS interior into
a pulsar-like exterior magnetosphere~\cite{prs15}. Here we survey different BHNS configurations
that differ in the initial BH spin, mass ratio, and magnetic field topology to study the
robustness of the jet launching scenario. Although the numerical studies reported here are
illustrative and not exhaustive, they suggest the following:

Varying the initial spin of the BH companion in the binary from $\tilde{a}=-0.5$ to $0.5$, we
observe that only the higher spin BHNS configuration launches a jet. In the antialigned case
Aliq3sm0.5, the star basically plunges into the black hole leaving a weakly magnetized matter
(less than $1\%$ of the initial rest-mass of the star) to form the disk (see~Table
\ref{table:results_allBHs}). We do not find any evidence of large-scale magnetic field collimation
or an outflow for this case. By contrast, in  Aliq3sp0.0 we did observe magnetic field collimation
above the BH poles, but after $t-t_{\rm GW}\sim 7000M\sim 175(M_{\rm NS}/1.4M_\odot)$ms the magnetic
pressure gradients were still too weak to launch an outflow. The lack of an outflow may be attributed
to the persistent fall-back toward the BH observed as we terminated the simulation. When the atmosphere
above the BH poles becomes thinner as the accretion proceeds, we anticipate that the magnetic pressure
may eventually overcome the ram pressure. However, jet launching may not be possible if the onset
time is longer than the lifetime of the accretion disk [$\tau_{\rm disk}\sim 0.36(M_{\rm NS}
  /1.4M_\odot)$s]. The mass of the disk is determined  by how far from the ISCO 
tidal disruption occurs. According, for a given NS companion,  the above results indicate that
there is a threshold value for the initial BH spin below which the jet launching cannot occur.

Varying the mass ratio of our BHNS configurations from $q=3:1$ to $q=5:1$, we
find that only remnants with  sizable accretion disks, and consequently considerable
magnetic energy, may launch a jet. Taking into account population  synthesis studies (see e.g.
\cite{Belczynski:2007xg,bdbofh10}) that suggest that the most likely BHNS mass ratio
may be $q =7:1$,~we estimated that the critical spin at which tidal disruption occurs at the ISCO is
$\tilde{a}=0.4$ (see also~\cite{Kyutoku:2011vz}). As the basic ingredient for jet launching is a
sizable magnetized disk, the above estimate suggests that high mass ratio BHNS systems can be the
central engines that power sGRBs only if the binary contains a highly spinning BH
($\tilde{a}\gtrsim 0.4$).

Finally, varying the direction of the magnetic field with respect to the total angular momentum
of the system from an aligned configuration  to a $90^\circ$-tilted configuration, we found that the
disk + BH remnant in the latter case lacks of a coherent poloidal magnetic field configuration. 
At after about  $t-t_{\rm GW}\sim 4000M\sim 100(M_{\rm NS}/1.4M_\odot) \rm ms$ we did not see any
indication of magnetic field collimation or an outflow.  A poloidal magnetic field component
with a consistent sign in the vertical direction is required to launch and support a jet
\cite{GRMHD_Jets_Req_Strong_Pol_fields}. These results suggest thus that there may also be a threshold
value of the tilt angle of the magnetic dipole moment above  which there are no jets.

A caveat is in order. Our GRMHD simulations do not account for all the physical processes
involved in BHNS mergers. In particular, it has been suggested that neutrino annihilation
in disk + BH systems may carry away a significant amount of energy from inner regions of
the accretion disks that may be strong enough to power jets~\cite{Popham:1998ab,Matteo:2002ck,
  Chen:2006rra,Lei2013ApJ,Just:2015dba}. Recently, it was suggested in~\cite{Lei:2017zro} that
the emergence of a jet in slowly BH + spinning disk systems may be triggered by neutrino-annihilation
and then by the BZ mechanism, leading to a transition from a thermally-dominated fireball to a
Poynting EM-dominated flow as is inferred for some GRBs, such as GRB 160625B~\cite{2018NatAs...2...69Z}.
We plan to study such processes in the future.
 

\acknowledgements
We thank V. Paschalidis for useful discussions, and the Illinois Relativity group REU
team (Eric Connelly, Kyle Nelli, and John Simone) for assistance with some of the
visualizations. This work has been supported in part by National Science Foundation
(NSF) Grant PHY-1602536 and PHY-1662211, and NASA Grant
80NSSC17K0070 at the University of Illinois at Urbana-Champaign.
This work made use of the Extreme Science and Engineering Discovery
Environment (XSEDE), which is supported by National Science Foundation
grant number TG-MCA99S008. This research is part of
the Blue Waters sustained-petascale computing project,
which is supported by the National Science Foundation
(awards OCI-0725070 and ACI-1238993) and the State of
Illinois. Blue Waters is a joint effort of the University
of Illinois at Urbana-Champaign and its National Center
for Supercomputing Applications.

\bibliographystyle{apsrev4-1}        
\bibliography{references}            
\end{document}